\documentclass[11pt]{article}

\usepackage[top=50mm, bottom=50mm, left=50mm, right=50mm]{geometry}

\usepackage{amsmath}
\usepackage{amssymb}
\usepackage{amsfonts}
\usepackage{mathtools}
\usepackage{mathrsfs}
\usepackage[scr = dutchcal]{mathalpha}
\usepackage{graphicx}
\usepackage{graphics}
\usepackage{float}
\usepackage{color}
\usepackage{fullpage}
\usepackage[normalem]{ulem} 
\usepackage{makeidx}
\usepackage{xspace}
\usepackage{authblk}
\usepackage{datetime}
\usepackage{microtype}
\usepackage{xpatch}
\usepackage{xstring}
\usepackage{fancyhdr}
\usepackage{mathrsfs}
\usepackage{esint}
\usepackage{upgreek}
\usepackage{siunitx}
\usepackage{chemformula}
\usepackage{tabularx, multirow}
\usepackage{threeparttable}
\usepackage[]{booktabs} 
\usepackage{float}
\usepackage[section]{placeins}
\usepackage{graphicx,xcolor}
\usepackage[colorlinks=false]{hyperref}
\usepackage[noabbrev,sort&compress]{cleveref}
\usepackage[font = footnotesize, labelfont=bf]{caption}
\usepackage{setspace}
\usepackage{cite}
\usepackage[square,numbers,sort&compress]{natbib}

\fancypagestyle{fpg}{%
	\lfoot{\footnotesize Preprint submitted to Materials \& Design }%
	\cfoot{\normalsize\thepage}
	\rfoot{\footnotesize \textbf{2022}\href{}{}}%
}
\newcommand{\doi}[1]{DOI: \href{http://dx.doi.org/#1}{\nolinkurl{#1}}}
\makeindex

\author[1,$\dagger$]{Amin Ebrahimi}
\author[2]{Mohammad Sattari}
\author[2]{Scholte J.L. Bremer}
\author[2]{Martin Luckabauer}
\author[2]{Gert-willem~R.B.E.~Römer}
\author[1]{Ian~M.~Richardson}
\author[3]{Chris~R.~Kleijn}
\author[1]{Marcel~J.M.~Hermans}

\affil[1]{\textit{Department of Materials Science and Engineering, Faculty of Mechanical, Maritime and Materials~Engineering, Delft~University~of~Technology, Mekelweg~2, 2628CD~Delft, The~Netherlands}}
\affil[2]{\textit{Department of Mechanics of Solids, Surfaces and Systems, Faculty of Engineering Technology, University~of~Twente, Drienerlolaan~5, 7522NB~Enschede, The~Netherlands}}
\affil[3]{\textit{Department of Chemical Engineering, Faculty of Applied Sciences, Delft~University~of~Technology, van~der~Maasweg~9, 2629HZ~Delft, The~Netherlands}}
\affil[$\dagger$]{Corresponding author: A.Ebrahimi@tudelft.nl (A. Ebrahimi)}

\title{\Large\textbf{The~influence of laser characteristics on internal flow behaviour in laser melting of metallic substrates}}
\date{}

\begin{document}

\maketitle
\thispagestyle{fpg}

\begin{abstract}
		The~absorptivity of a~material is a~major uncertainty in numerical simulations of laser welding and additive manufacturing, and its value is often calibrated through trial-and-error exercises. This adversely affects the~capability of numerical simulations when predicting the~process behaviour and can eventually hinder the~exploitation of fully digitised manufacturing processes, which is a~goal of ``industry~4.0". In the~present work, an~enhanced absorption model that takes into account the~effects of laser characteristics, incident angle, surface temperature, and material composition is utilised to predict internal heat and fluid flow in laser melting of stainless steel 316L. Employing such an~absorption model is physically more realistic than assuming a~constant absorptivity and can reduce the~costs associated with calibrating an~appropriate value. High-fidelity three-dimensional numerical simulations were performed using both variable and constant absorptivity models and the~predictions compared with experimental data. The~results of the~present work unravel the crucial effect of absorptivity on the~physics of internal flow in laser material processing. The~difference between melt-pool shapes obtained using fibre and \ch{CO2} laser sources is explained, and factors affecting the~local energy absorption are discussed.
\end{abstract}

\bigskip
\newpage

\onehalfspacing
\section{Introduction}
\label{sec:intro}
Laser-beam melting of metallic substrates forms the~basis of many advanced fusion-based manufacturing processes (such as laser welding, laser cladding, laser metal deposition (LMD), and selective laser melting (SLM)) and has brought new perspectives on advancement of materials processing and manufacturing of high-integrity products. Successful adoption of laser-beam melting in real-world engineering applications requires finding processing windows within which the~product quality should meet the~intended standards~\cite{Khairallah_2020}. However, determining the~processing window through trial-and-error experiments is challenging and involves significant costs due to the~large number of process parameters and the~coupling between various physical phenomena. Simulation-based approaches have been recognised as a~promising alternative to costly and time-inefficient experiments and can be utilised to reduce the~costs of design-space exploration~\cite{Markl_2016,Ebrahimi_2021}. Moreover, numerical simulations can enhance our understanding of the~complex transport phenomena in laser material processing that are not easily accessible through experiments~\cite{Francois_2017,DebRoy_2020}.

Successful adoption of simulation-based approaches for process development and optimisation relies predominantly on adequate modelling of various physical phenomena that occur during laser melting (\textit{e.g.}~laser-matter interaction, heat and fluid flow, and solid-liquid phase transformation)~\cite{Cook_2020}. Assumptions made to develop computational models often affect their reliability, accuracy and performance in predicting and describing the~process behaviour. For instance, studies suggest that melt-pool surface deformations affect power-density distribution, leading to changes in the~thermal field, Marangoni flow pattern and the~melt-pool shape~\cite{Simonds_2021,Ebrahimi_2020,Ebrahimi_2021_b}. {Conversely, previous investigations~\cite{Xie_1997,Mahrle_2009,Ren_2021,Yang_2021} have shown a~considerable influence of laser characteristics and power-density distribution on molten metal flow behaviour in laser welding and additive manufacturing. Thus, there seems to be an~important bi-directional coupling between laser power-density distribution and melt-pool behaviour.} Neglecting such effects in numerical simulations of laser-beam melting can negatively affect the~quality of numerical predictions of thermal fields, microstructures and properties of the~product~\cite{Ebrahimi_2020,Shu_2021}. Moreover, assumptions made to develop a~computational model may necessitate the~incorporation of unphysical tuning parameters to obtain agreement between numerical and experimental data~\cite{Ebrahimi_2020,Kidess_2016_thermalSci}. This can reduce the~model reliability for design-space explorations since a~change in process parameters or material properties may require recalibrating the~tuning parameters~\cite{Grange_2021,De_2004}. Understanding the~influence of such assumptions on numerical predictions is therefore essential and can guide the~modelling efforts to enhance the~current numerical simulations.

Absorption of laser-energy, energy-density distribution and its variation over time are critical components influencing the~modelling of laser-beam melting~\cite{King_2015} and depend on a~variety of process parameters including the~characteristics of the~laser system (\textit{e.g.}~laser intensity and wavelength), thermophysical properties of the~material, surface roughness and chemistry, and interactions of the~melt-pool surface with the~laser beam~\cite{Khairallah_2014,Indhu_2018,Ye_2019,Svetlizky_2021}. In the~majority of previous studies on laser-beam melting, the~absorption of laser energy is assumed to be constant~\cite{Yang_2021}, neglecting the~unsteady interactions between laser-beam and material surface~\cite{Trapp_2017}. Studies have shown that changes in melt-pool surface morphology and temperature can affect the~local absorptivity of the~material~\cite{Ready_1997,Xie_1997,Katayama_2013,Kouraytem_2019}. The~Fresnel absorption model~\cite{Lvovsky_2015}, which is commonly employed in numerical simulations of laser melting (particularly when the~ray-tracing method is used), accounts for the~effects of laser-ray incident angle and material refractive index, but neglects the~temperature dependence of material absorptivity~\cite{Indhu_2018}. In laser melting of metallic substrates, the~material often experiences large changes in temperature that can significantly affect the~thermophysical properties of the~material, including the~material absorptivity~\cite{Bass_1983,Simonds_2018}. Moreover, the~complex molten metal flow in melt-pools continuously disturbs temperature distribution over the~surface~\cite{Ebrahimi_2020}, affecting the~local absorptivity of the~material. The~Fresnel model cannot reflect the~variation of local energy absorption that occurs due to changes in melt-pool surface temperature~\cite{Ujihara_1972,Wang_2021,Ren_2021}. Hence, the~Fresnel model cannot describe variations in material absorptivity with sufficient accuracy, particularly in cases where the~melt-pool surface deformations are small compared to the~melt-pool depth (for instance, laser cladding, conduction-mode laser welding and laser metal deposition).

{Realising that in practical applications the~laser type is not a~control parameter, as commercial machines come with a~fixed laser type, the~present work focuses on understanding the~influence of laser characteristics on complex heat and molten metal flow in laser-beam melting. Such an~understanding allows us to explain, for example, the~difference between melt-pool shapes obtained using fibre transmissible and \ch{CO2} lasers (\textit{i.e.} the most widely employed lasers for industrial applications).}
High-fidelity three-dimensional numerical simulations are performed using an~enhanced~laser-beam absorption model that takes into account the~effects of laser characteristics, surface temperature, incident angle and base-material composition. The~results obtained using the~enhanced absorption model for different laser systems and laser powers are compared with those obtained using a~constant absorptivity and factors affecting the~local energy absorption are discussed. Additionally, experiments are carried out for different laser powers to validate the~melt-pool shapes predicted using the~present computational model. The~results and discussions provided in the~present work guide the~modelling efforts to improve simulations of fusion-based welding and additive manufacturing.

\section{Problem description}
\label{sec:problem_des}
\begin{figure}[!htb] 
	\centering
	\includegraphics[width=0.9\linewidth]{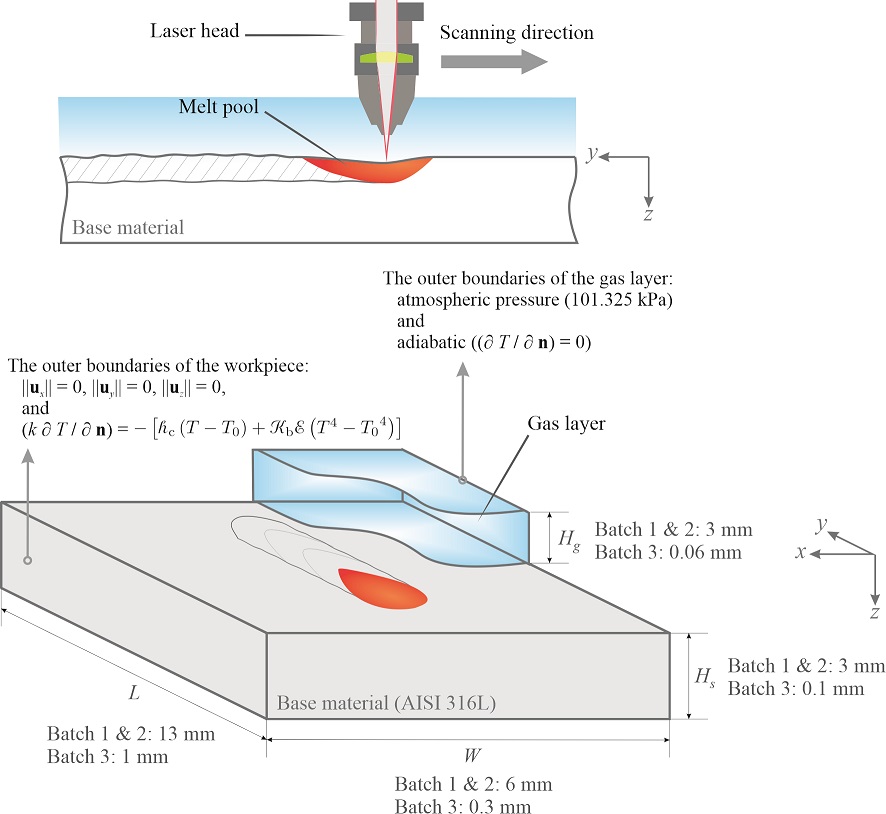}
	\caption{Schematic of laser melting, dimensions of the~computational domain and the~mathematical expressions used for the~boundary conditions. Parts of the~gas layer are clipped for visualisation.}
	\label{fig:schematic}
\end{figure}

As shown in \cref{fig:schematic}, a~moving laser beam is employed to locally heat and melt the~substrate that is made of a~stainless steel alloy (AISI~316L) and is initially at an~ambient temperature of~$\SI{300}{\kelvin}$. The~gas layer above the~plate is included in simulations to track the~motion of gas-metal interface. The~influence of laser characteristics on melting of a~metallic substrate is studied numerically for fibre and \ch{CO2} lasers, whose wavelengths are different. The~laser beam is perpendicular to the~substrate surface and has a~Gaussian intensity profile. 

\begin{figure}[!htb] 
	\centering
	\includegraphics[width=1.0\linewidth]{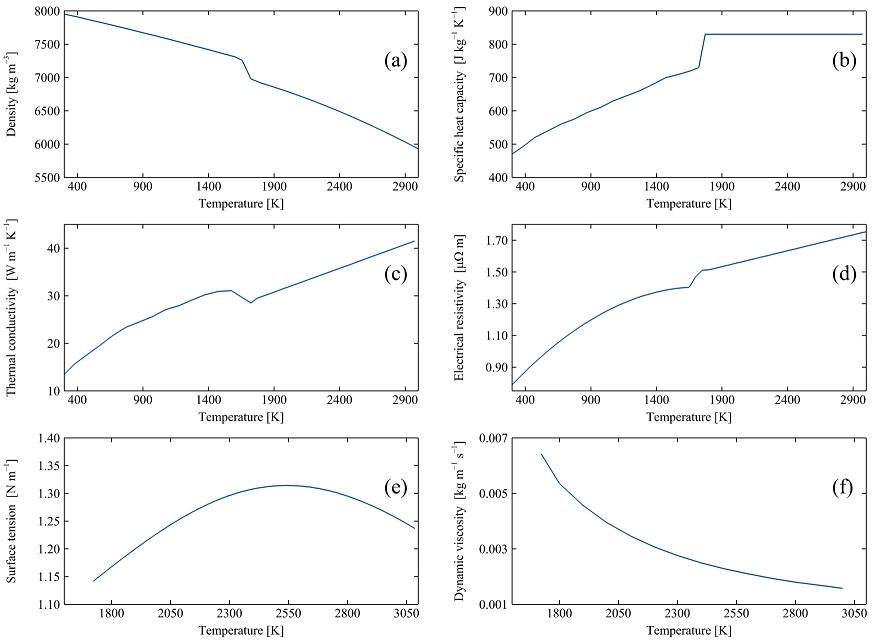}
	\caption{Temperature-dependent properties of stainless steel 316L. (a)~density~\cite{Kim_1975}, (b)~specific heat capacity~\cite{Mills_2002_316}, (c)~thermal conductivity~\cite{Mills_2002_316}, (d)~electrical resistivity~\cite{Pichler_2019} (e)~surface~tension~\cite{Sahoo_1988} and (f)~dynamic viscosity~\cite{Kim_1975}.}
	\label{fig:materials_properties}
\end{figure}

Three batches of three-dimensional numerical simulations are executed for different laser types and powers using both constant and variable absorptivity models to describe the~complex thermal and fluid flow fields in the~melt pool. For cases where the~absorptivity is assumed to be constant, different values of the~absorptivity are examined, as reported in~\cref{tab:simulations}. The~dimensions of the~computational domain defined in a~Cartesian coordinate system, and the~boundary conditions applied to the~outer boundaries, are shown in~\cref{fig:schematic}. Heat input from the~laser, heat losses due to convection, radiation and vaporisation, as well as forces acting on the~melt pool (\textit{i.e.}~Marangoni shear force, capillary force, and recoil pressure) are implemented in the~simulations by adding source and sink terms to the~governing equations, as described in~\cref{sec:methods}. Although the~thermal buoyancy force is often negligible compared with Marangoni force in driving the~molten metal flow in laser melting~\cite{Ebrahimi_2020}, a~variable density model is employed in the~present work to account for thermal buoyancy force as well as the~solidification shrinkage. Temperature-dependent material properties are employed for both the~solid and the~molten metal in the~present numerical simulations and the~values are reported in \cref{tab:material_properties} and \cref{fig:materials_properties}. Although the~properties of argon are also temperature-dependent, they have been assumed to be constant in the~present work for the~sake of simplicity, see \cref{tab:material_properties}. This assumption is made based on the~fact that the~density, viscosity and thermal conductivity of argon are very small compared to those of the~metal, therefore variations of those gas properties with temperature have negligible influence on the~numerical predictions~\cite{Saldi_2012_thesis}. 

\bgroup
\def\arraystretch{1.15}	
\begin{table}[!htb] 
	\centering
	\begin{threeparttable}[t]
		\caption{Summary of the~process parameters studied in the~present work.}
		\small
		\begin{tabular}{lllll}
			\toprule
			Parameter                                                                   & Batch 1                    & Batch 2        & Batch 3               &  \\ \midrule
			Laser type                                                                  & \ch{CO2} laser             & Nd:YAG laser   & Fibre laser \tnote{a} &  \\
			Laser power~$\mathscr{P}$~[\si{\watt}]                                      & 900--2100 (interval: 300)  & 500--900 (100) & 200                   &  \\
			Wavelength~$\lambda$~[\si{\meter}]                                          & \SI{1.060e-5}              & \SI{1.064e-6}  & \SI{1.070e-6}         &  \\
			Constant absorptivity~$\mathscr{a}$~[--]                                    & 0.10--0.14 (0.02) and 0.18 & 0.3--0.4 (0.05)   & 0.3--0.4 (0.05)          &  \\
			Travel speed~$\mathscr{V}$~[\si{\meter\per\second}]                         & $10^{-2}$                  & $10^{-2}$      & \SI{1.5}              &  \\
			Spot size~(D4$\sigma$)~$d_\mathrm{b}$~[\si{\meter}]                         & \SI{2e-3}                  & \SI{2e-3}      & \SI{1.1e-4}           &  \\
			Interaction time~$t_\mathrm{i} = d_\mathrm{b} / \mathscr{V}$~[\si{\second}] & \SI{0.2}                   & \SI{0.2}       & \SI{7.3e-5}           &  \\ \bottomrule
			&                            &                &                       &
		\end{tabular} 
		\label{tab:simulations}
		\begin{tablenotes}
			\footnotesize{\item[a] continuous wave fibre laser (YLR-500-AC, IPG Photonics)~\cite{Khairallah_2020}}
		\end{tablenotes}
	\end{threeparttable}
\end{table}
\egroup	

\bgroup
\def\arraystretch{1.0}	
\begin{table}[!htb] 
	\centering
	\caption{Material properties employed in the~present work.}
	\small
	\begin{tabular}{llll}
		\toprule
		Property                                                                        & Stainless steel (AISI~316L)                          & {Gas (argon)~\cite{Jaques_1988}} &  \\ \midrule
		Density $\rho$ [\si{\kilogram\per\meter\cubed}]                                 & see \cref{fig:materials_properties}                  & \SI{1.623}                                       &  \\
		Specific heat capacity $c_\mathrm{p}$ [\si{\joule\per\kilogram\per\kelvin}]     & see \cref{fig:materials_properties}                  & \SI{520.64}                                      &  \\
		Thermal conductivity $k$ [\si{\watt\per\meter\per\kelvin}]                      & see \cref{fig:materials_properties}                  & \SI{1.58e-2}                                     &  \\
		Dynamic viscosity $\mu$ [\si{\kilogram\per\meter\per\second}]                   & see \cref{fig:materials_properties}                  & \SI{2.12e-05}                                    &  \\
		Molar mass  $M$ [\si{\kilogram\per\mole}]                                       & {\SI{5.58e-2}~\cite{Mills_2002_316}} & \SI{3.9948e-2}                                   &  \\
		Latent heat of fusion $\mathcal{L}_\mathrm{f}$ [\si{\joule\per\kilogram}]       & {\SI{2.7e5}~\cite{Mills_2002_316}}   & --                                               &  \\
		Latent heat of vaporisation $\mathcal{L}_\mathrm{v}$ [\si{\joule\per\kilogram}] & {\SI{7.45e6}~\cite{Heeling_2017}}    & --                                               &  \\
		Solidus temperature $T_\mathrm{s}$ [\si{\kelvin}]                               & {\SI{1658}~\cite{Mills_2002_316}}    & --                                               &  \\
		Liquidus temperature $T_\mathrm{l}$ [\si{\kelvin}]                              & {\SI{1723}~\cite{Mills_2002_316}}    & --                                               &  \\
		Boiling temperature $T_\mathrm{b}$ [\si{\kelvin}]                               & {\SI{3086}~\cite{Khairallah_2016}}   & --                                               &  \\ \bottomrule
	\end{tabular} 
	\label{tab:material_properties}
\end{table}
\egroup	

\FloatBarrier

\section{Methods} 
\label{sec:methods}

\subsection{Model formulation}
\label{sec:math_model}	
The~present computational model is developed on the~basis of the~finite-volume approach and utilises the~volume-of-fluid (VOF) method~\cite{Hirt_1981} to locate the~gas-metal interface. It is assumed that both the~molten metal and argon are Newtonian fluids and that their densities are pressure-independent. Based on these assumptions, the~governing equations for conservation of mass, momentum and energy as well as for the~advection of the~scalar function~$\phi$, which represents the~local volume-fraction of the~steel phase in a~computational cell, are defined as follows:

\begin{equation}
	\frac{D \rho}{D t} + \rho \left(\nabla \cdot \mathbf{u}\right) = 0,
	\label{eq:mass}
\end{equation}

\begin{equation}
	\rho \frac{D \mathbf{u}}{D t} = \mu \nabla^2 \mathbf{u} -\nabla p + \mathbf{F}_\mathrm{d} + \mathbf{F}_\mathrm{s},
	\label{eq:momentum}
\end{equation}

\begin{equation}
	\rho \frac{D h}{D t} = \frac{k}{c_\mathrm{p}} \nabla^2 h - \rho \frac{D \left( \psi \mathcal{L}_\mathrm{f} \right)}{D t} + S_\mathrm{q} + S_\mathrm{l},
	\label{eq:energy}
\end{equation}

\begin{equation}
	\frac{D \phi}{D t} = 0.
	\label{eq:vof}
\end{equation}

\noindent
Here, $\rho$ is the~density, $t$~the~time, $\mathbf{u}$~the~fluid velocity vector, $\mu$~the~dynamic viscosity, $p$~the~pressure, $h$~the~sensible heat, $k$~the~thermal conductivity, $c_\mathrm{p}$~the~specific heat capacity, and $\left( \psi \mathcal{L}_\mathrm{f} \right)$~the~latent~heat with $\psi$~being the~local liquid volume-fraction and $\mathcal{L}_\mathrm{f}$~the~latent heat of fusion. It is assumed that the~liquid volume-fraction~$\psi$ varies linearly with temperature~\cite{Voller_1991}, and its value is calculated as follows:

\begin{equation}
	\psi = \frac{T - T_\mathrm{s}}{T_\mathrm{l} - T_\mathrm{s}}; \quad T_\mathrm{s} \le T \le T_\mathrm{l}.
	\label{eq:liquid_fraction}
\end{equation}

\noindent
The~effective material properties in computational cells were computed as follows:

\begin{equation}
	\xi = \phi \, \xi_\mathrm{m} + \left(1-\phi\right) \xi_\mathrm{g},
	\label{eq:mixture_model}
\end{equation}

\noindent
where,~$\xi$ corresponds to density~$\rho$, specific heat capacity~$c_\mathrm{p}$, thermal conductivity~$k$ or viscosity~$\mu$, and subscripts `m' and `g' indicate metal or gas respectively.

$\mathbf{F}_\mathrm{d}$ is a~sink term incorporated into \cref{eq:momentum} to damp fluid velocities in the~mushy region and to suppress them in solid regions, and is defined based on the~enthalpy-porosity technique~\cite{Voller_1987} as

\begin{equation}
	\mathbf{F}_\mathrm{d} = -C\ \frac{(1 - \psi)^2}{\psi^3 + \epsilon} \ \mathbf{u},
	\label{eq:sink_term}
\end{equation}

\noindent
where,~$C$ is the~permeability coefficient (also known as the~mushy-zone constant) and~$\epsilon$ is a~constant, equal to~$10^{-3}$, employed to avoid division by zero. The~value of the~permeability coefficient~$C$ was set to $10^7\,\SI{}{\kilogram\per\square\meter\per\square\second}$, based on the~criteria proposed by Ebrahimi~\textit{et al.}~\cite{Ebrahimi_2019}.

A~continuum surface force model~\cite{Brackbill_1992} is used to model forces acting on the~gas-metal interface (\textit{i.e.}~capillary force, Marangoni shear force and recoil pressure). Accordingly, the~source term~$\mathbf{F}_\mathrm{s}$ is incorporated into \cref{eq:momentum} as follows:

\begin{equation}
	\mathbf{F}_\mathrm{s} = \mathbf{f}_\mathrm{s} \lVert \nabla\phi \rVert \frac{2\rho}{\rho_\mathrm{m} + \rho_\mathrm{g}},
	\label{eq:CSF_model}
\end{equation}

\noindent
where, $\mathbf{f}_\mathrm{s}$ is the~surface force applied to a~unit area, and the~term~$2\rho/\!\left(\rho_\mathrm{m} + \rho_\mathrm{g}\right)$ is included to redistribute the~surface-forces towards the~heavier phase. The~surface force $\mathbf{f}_\mathrm{s}$ is determined as follows:

\begin{equation}
	\begin{split}
		\mathbf{f}_\mathrm{s} &= \mathbf{f}_\mathrm{capillary} + \mathbf{f}_\mathrm{Marangoni} + \mathbf{f}_\mathrm{P_{recoil}} \\
		& = \gamma \kappa \hat{\mathbf{n}} + \frac{\mathrm{d} \gamma}{\mathrm{d} T} \left[\nabla T - \hat{\mathbf{n}}\left(\hat{\mathbf{n}} \cdot \nabla T\right)\right] + \left[0.54 \cdot p_0 \exp\left(\frac{\mathcal{L}_\mathrm{v} \, M \, \left(T - T_\mathrm{b}\right)}{\mathrm{R} \, T \, T_\mathrm{b}}\right)\right] \hat{\mathbf{n}}.
	\end{split}
	\label{eq:surface_force}
\end{equation}

\noindent
where, $\gamma$~is the~surface tension, $\hat{\mathbf{n}}$~the~surface unit normal vector~\mbox{($\hat{\mathbf{n}} = \nabla\phi / \lVert \nabla\phi \rVert$)}, $\kappa$ the~surface curvature~($\kappa = \nabla\cdot\hat{\mathbf{n}}$), $p_0$~the~ambient pressure, and $\mathrm{R}$ the~universal gas constant. {The~third term on the~right-hand side of~\cref{eq:surface_force}, $\mathbf{f}_\mathrm{P_{recoil}}$, is included to model the~recoil pressure generated due to vaporisation of the~molten metal~\cite{Anisimov_1995,Lee_2002}.}

The~source~$S_\mathrm{q}$ and the~sink term~$S_\mathrm{l}$ are incorporated into \cref{eq:energy} to model the~laser heat input to the~material and heat losses from the~material due to convection, radiation and vaporisation respectively, and are defined as follows:

\begin{equation}
	S_\mathrm{q} = \mathscr{F}_\mathrm{q} \left[\frac{2\, \mathscr{a} \, \mathscr{P}}{\pi r_\mathrm{b}^2} \exp\left(\frac{-2\, \mathscr{R}^2}{r_\mathrm{b}^2}\right) \lVert \nabla\phi \rVert \frac{2\, \rho \, c_\mathrm{p}}{(\rho \, c_\mathrm{p})_\mathrm{m} + (\rho \, c_\mathrm{p})_\mathrm{g}} \right],
	\label{eq:heat_source}
\end{equation}

\begin{equation}
	S_\mathrm{l} = - \left(S_\mathrm{convection} + S_\mathrm{radiation} + S_\mathrm{vaporisation} \right) \lVert \nabla\phi \rVert \frac{2\, \rho \, c_\mathrm{p}}{(\rho \, c_\mathrm{p})_\mathrm{m} + (\rho \, c_\mathrm{p})_\mathrm{g}}.
	\label{eq:heat_loss}
\end{equation}

\noindent
Here, $\mathscr{a}$~is the~absorptivity, $\mathscr{P}$~the~laser power, $r_\mathrm{b}$~the~radius of the~laser beam, $\mathscr{R}$~the~radial distance from the~laser-beam axis in~$x$-$y$ plane, and

\begin{equation}
	S_\mathrm{convection} = \mathscr{h}_\mathrm{c} \left(T - {T_\mathrm{0}}\right),
	\label{eq:heat_loss_convection}
\end{equation}

\begin{equation}
	S_\mathrm{radiation} =  \mathscr{K}_\mathrm{b} \mathscr{E} \left(T^4 - {T_\mathrm{0}}^4\right),
	\label{eq:heat_loss_radiation}
\end{equation}

\begin{equation}
	S_\mathrm{vaporisation} = 0.82 \cdot \frac{\mathcal{L}_\mathrm{v} \, M}{\sqrt{2 \pi M \, \mathrm{R} \, T}}\, p_0 \exp\left(\frac{\mathcal{L}_\mathrm{v} \, M \, \left(T - T_\mathrm{b}\right)}{\mathrm{R} \, T \, T_\mathrm{b}}\right),
	\label{eq:heat_loss_vaporisation}
\end{equation}

\noindent
where, $T_\mathrm{0}$ is the~ambient temperature, $\mathscr{K}_\mathrm{b}$ the~Stefan-Boltzmann constant, and $\mathscr{h}_\mathrm{c}$ and $\mathscr{E}$ are the~heat transfer coefficient and the~radiation emissivity equal to $\SI{25}{\watt\per\square\meter\per\kelvin}$~\cite{Johnson_2017} and 0.45~\cite{Sridharan_2011} respectively. Compared to the~total laser energy absorbed by the~material, the~heat losses from the~material due to convection and radiation are quite small; thus, the~precise values of $\mathscr{h}_\mathrm{c}$ and $\mathscr{E}$ are not critical in the~present simulations. {The~coefficient 0.82 in~\cref{eq:heat_loss_vaporisation} is included based on Anisimov's theory~\cite{Anisimov_1995} to account for the~reduced cooling effect due to metal vapour condensation.}

In the~VOF method, the~energy fluxes applied to the~material surface are included as volumetric source terms in the~computational cells that encompass the~melt-pool surface (\textit{i.e.}~cells with $0 < \phi < 1$). Hence, melt-pool surface deformations that occur during the~process can result in an~increase in the~total heat input to the~material~\cite{Ebrahimi_2020}. $\mathscr{F}_\mathrm{q}$ in \cref{eq:heat_source} is a~dynamic adjustment factor introduced to abate artificial increase in energy absorption due to surface deformations and is defined as

\begin{equation}
	\mathscr{F}_\mathrm{q} = \frac{1}{\iiint \limits_{\mathrm{\forall}} \lVert \nabla\phi \rVert \frac{2\, \rho \, c_\mathrm{p}}{(\rho \, c_\mathrm{p})_\mathrm{m} + (\rho \, c_\mathrm{p})_\mathrm{g}} \mathop{}\!\mathrm{d}V},
	\label{eq:adjustment_coefficient}
\end{equation}

\noindent
where, ``$\forall$" indicates the~computational domain. The~influence of utilising such an~adjustment factor on numerical predictions of molten metal flow in laser melting is explained by~Ebrahimi~\textit{et al.}~\cite{Ebrahimi_2020}.

\subsubsection{Absorptivity model}
\label{sec:absorptivity_model}	
When a~laser beam with total energy of~$\mathscr{P}$ interacts with material, here stainless steel 316L, part of its energy is absorbed by the~material for a~fraction equal to its absorptivity~($\mathscr{a}$). Numerical simulations developed for laser welding and additive manufacturing commonly assume the~absorptivity to be constant~\cite{Khairallah_2016}, which is physically unrealistic, and its value is often regarded as a~calibration parameter~\cite{Yang_2021}. The~absorptivity should be considered as a~material property and not a~calibration parameter~\cite{Yang_2021}. In the~present work, the~amount of laser energy absorbed by the~material was modelled according to the~absorptivity model proposed by~\mbox{Yang~\textit{et al.}~\cite{Yang_2021}} and Mahrle~and~Beyer~\cite{Mahrle_2009}, which takes into account the~effects of laser characteristics, laser-ray incident angle, surface temperature and base-material composition. {Accordingly, the~absorptivity $\mathscr{a}$ for circularly polarised or un-polarised laser radiation was approximated as follows~\cite{Ducharme_1994}:}

\begin{equation}
	\mathscr{a} = 1 - \frac{R_\mathrm{s} + R_\mathrm{p}}{2},
	\label{eq:absorptivity}
\end{equation}

\noindent
{where, according to the~Fresnel’s reflection equations, $R_\mathrm{s}$ and $R_\mathrm{p}$ are the~reflectance for parallel and perpendicularly polarised light~\cite{Katayama_2013} defined as}

\begin{equation}
	R_\mathrm{s} = \frac{\alpha^2 + \beta^2 - 2\alpha \cos(\theta) + \cos^2(\theta)}{\alpha^2 + \beta^2 + 2\alpha \cos(\theta) + \cos^2(\theta)},
	\label{eq:rs}
\end{equation}

\begin{equation}
	R_\mathrm{p} = R_\mathrm{s}\left(\frac{\alpha^2 + \beta^2 - 2\alpha \sin(\theta) \tan(\theta) + \sin^2(\theta)\tan^2(\theta)}{\alpha^2 + \beta^2 + 2\alpha \sin(\theta) \tan(\theta) + \sin^2(\theta)\tan^2(\theta)}\right).
	\label{eq:rp}
\end{equation}

\noindent
Here, $\theta$ is the~incident angle of the~laser ray, and $\alpha$ and $\beta$ are functions of the~refractive index $n$ and the~extinction coefficient $k$ of the~irradiated material. {The~values of $\alpha$, $\beta$, $n$ and $k$ were determined as follows~\cite{Yang_2021}:}

\begin{equation}
	\alpha = \left(\frac{\sqrt{\left(n^2 - k^2 - \sin^2(\theta)\right)^2 + 4n^2k^2} + n^2 - k^2 - \sin^2(\theta)}{2}\right)^{\frac{1}{2}},
	\label{eq:alpha}
\end{equation}

\begin{equation}
	\beta = \left(\frac{\sqrt{\left(n^2 - k^2 - \sin^2(\theta)\right)^2 + 4n^2k^2} - n^2 + k^2 + \sin^2(\theta)}{2}\right)^{\frac{1}{2}},
	\label{eq:beta}
\end{equation}

\begin{equation}
	n = \left(\frac{\sqrt{\mathscr{e}_\mathrm{r}^2 + \mathscr{e}_\mathrm{i}^2} + \mathscr{e}_\mathrm{r}}{2}\right)^{\frac{1}{2}},
	\label{eq:n}
\end{equation}

\begin{equation}
	k = \left(\frac{\sqrt{\mathscr{e}_\mathrm{r}^2 + \mathscr{e}_\mathrm{i}^2} - \mathscr{e}_\mathrm{r}}{2}\right)^{\frac{1}{2}},
	\label{eq:k}
\end{equation}

\noindent
{where, $\mathscr{e}_\mathrm{r}$ and $\mathscr{e}_\mathrm{i}$ are the~real and imaginary parts of the~relative electric permittivity~$\tilde{\mathscr{e}}$ respectively~\cite{Wooten_1972}}, defined as

\begin{equation}
	\mathscr{e}_\mathrm{r} = 1 - \frac{\omega_\mathrm{p}^2}{\mathscr{f}^2 + \delta^2},
	\label{eq:er}
\end{equation}

\begin{equation}
	\mathscr{e}_\mathrm{i} = \frac{\delta}{\mathscr{f}} \frac{\omega_\mathrm{p}^2}{\mathscr{f}^2 + \delta^2}.
	\label{eq:ei}
\end{equation}

\begin{figure}[!htb] 
	\centering
	\includegraphics[width=1.0\linewidth]{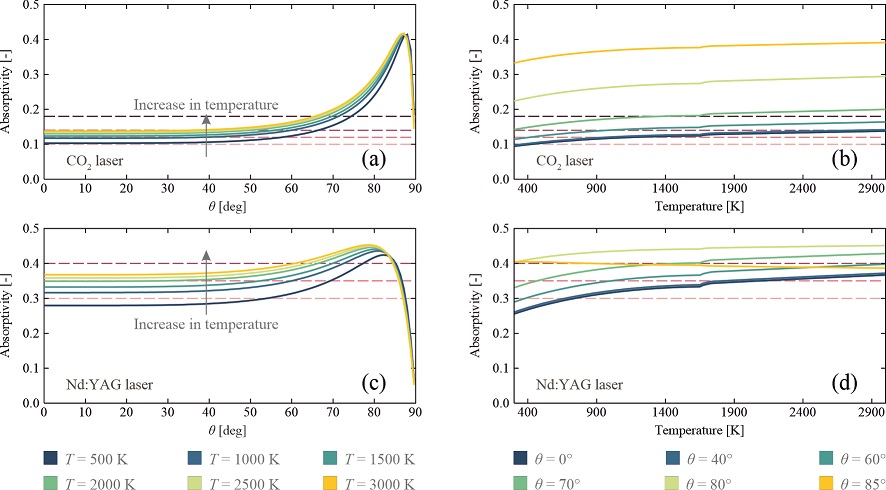}
	\caption{The~effects of laser emission wavelength $\lambda$, temperature and incident angle of the~laser ray $\theta$ on the~absorptivity of stainless steel 316L. Values are obtained from the~absorptivity model described in \cref{sec:absorptivity_model}. (a and b) \ch{CO2}~laser with $\lambda = \SI{1.060e-5}{\meter}$ and (c and d) Nd:YAG laser with $\lambda = \SI{1.064e-6}{\meter}$. Dashed lines indicate the~constant absorptivities studied in the~present work.}
	\label{fig:absorptivity}
\end{figure}

\noindent
{Here, $\omega_\mathrm{p}$, $\mathscr{f}$ and $\delta$ are the~plasma frequency, laser frequency and damping frequency respectively, defined as follows~\cite{Mahrle_2009,Yang_2021}:}

\begin{equation}
	\omega_\mathrm{p} = \sqrt{\frac{N_\mathrm{e} \, \mathrm{q}_\mathrm{e}^2}{\mathrm{M}_\mathrm{e} \, \mathrm{v}_0}},
	\label{eq:plasma_frequncy}
\end{equation}

\begin{equation}
	\mathscr{f} = \frac{2 \pi \ \mathscr{c}}{\lambda},
	\label{eq:laser_frequency}
\end{equation}

\begin{equation}
	\delta = \omega_\mathrm{p}^2 \, \mathrm{v}_0 \, \mathcal{R}_T,
	\label{eq:damping_frequency}
\end{equation}

\noindent
where, $N_\mathrm{e}$ is the~mean number density of free electrons (the~value of which approximately equals to $\SI{5.83e29}{\per\cubic\meter}$ for stainless steel 316L~\cite{Yang_2021}), $\mathrm{q}_\mathrm{e}$ the~elementary electric charge, $\mathrm{M}_\mathrm{e}$ the~electron rest mass, $\mathrm{v}_0$ the~electric constant, $\mathscr{c}$ the~speed of light in vacuum, $\lambda$ the~emission wavelength of the~laser, and $\mathcal{R}_T$ the~temperature-dependent electrical resistivity. The~values approximated using the~present absorptivity model for stainless steel 316L are presented in \cref{fig:absorptivity}, showing the~influence of laser emission wavelength, incident angle and temperature on the~absorptivity. The~data shown in \cref{fig:absorptivity} suggest that the~incident angle of the~laser ray $\theta$ negligibly affects the~absorptivity up to 40$^\circ$ but its effect becomes significant for larger incident angles (for instance, in cases where a~keyhole is formed). Additionally, changes in the~temperature notably affect absorptivity, making the~assumption of constant absorptivity questionable.

\FloatBarrier

\subsection{Numerical implementation}
\label{sec:num_proc}
\begin{figure}[!htb] 
	\centering
	\includegraphics[width=0.9\linewidth]{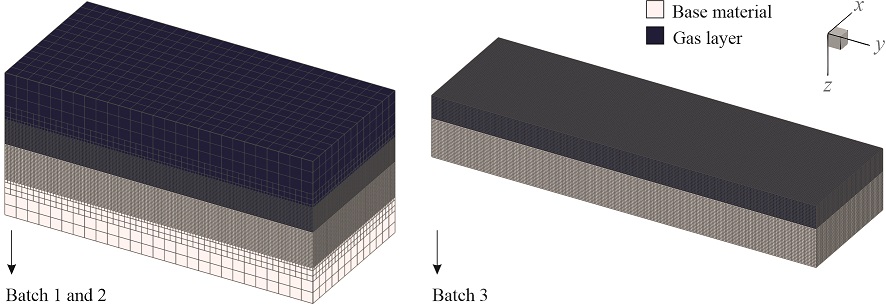}
	\caption{The~computational meshes employed in the~present work.}
	\label{fig:mesh}
\end{figure}

The~present numerical simulations were constructed on the~foundation of a~proprietary flow solver, ANSYS~Fluent~\cite{Ansys_192}. User-defined functions (UDFs) programmed in the~C programming language were developed to implement the~absorptivity model, source and sink terms in the~momentum and energy equations as well as the~surface tension model in the~simulations. {As shown in our previous works~\cite{Ebrahimi_2019_conf,Ebrahimi_2020,Ebrahimi_2021,Ebrahimi_2021_b,Ebrahimi_2021_c}, the~numerical grid cell spacing should be chosen to have at least 35 cells in the~melt pool region along its width. Accordingly, hexahedral cells were used to discretise the~computational domain with minimum cell spacing of $\SI{50}{\micro\meter}$ for cases in batch 1~and~2 (\ch{CO2} and Nd:YAG welds, melt-pool widths of about~\SI{1500}{\micro\meter}), and $\SI{3}{\micro\meter}$ for cases in batch 3 (continuous wave fibre laser welds, melt-pool widths of about~\SI{100}{\micro\meter}), as shown in \cref{fig:mesh}. } Accordingly, the~total number of computational cells is about~$\SI{1.2e6}{}$ for cases in batch 1 and 2, and about~$\SI{1.0e6}{}$ for cases in batch 3.

The~central differencing scheme with second-order accuracy and a~first-order implicit scheme were employed for spatial discretisation and time marching respectively. A~fixed time-step size $\Delta t$ was used in the~simulations and its value was chosen sufficiently small ($\SI[parse-numbers=false]{10^{-8}}{\second} < \Delta t < \SI[parse-numbers=false]{10^{-5}}{\second}$) to achieve a~Courant number $(\mathrm{Co} = \lVert \mathbf{u} \rVert \Delta t / \Delta x)$ less than $0.2$. The~PRESTO~(pressure staggering option) scheme~\cite{Patankar_1980} was used for the~pressure interpolation, and the~PISO~(pressure-implicit with splitting of operators) scheme~\cite{Issa_1986} was used to couple velocity and pressure fields. An~explicit compressive VOF method~\cite{Ubbink_1997} was employed to formulate the~advection of the~scalar field $\phi$. Each simulation was run in parallel on 16~cores~(AMD~EPYC~7452) of a~high-performance computing cluster with $\SI{256}{\giga B}$~memory.

\FloatBarrier

\subsection{Experimental setup and procedure}
\label{sec:experimental_setup}
\begin{figure}[!htb] 
	\centering
	\includegraphics[width=1.0\linewidth]{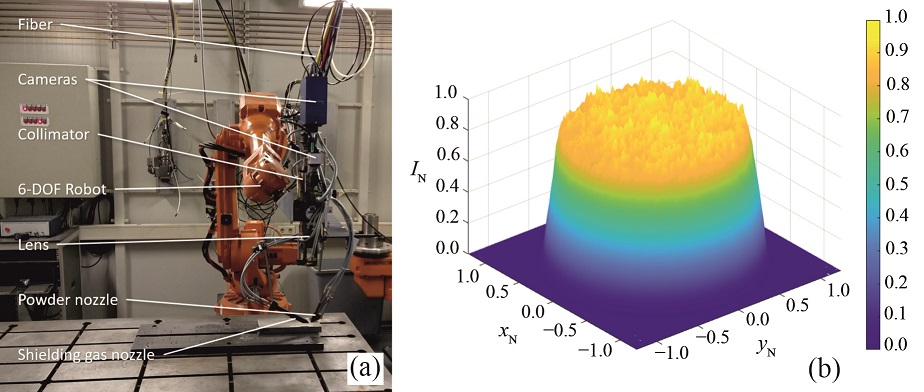}
	\caption{(a) The~experimental setup employed in the~present work. (b) Power-density distribution produced using the~present experimental setup. In subfigure (b), the~power-density profile is normalised with the~respective maximum peak, and coordinates are non-dimensionalised using the~laser-beam radius $r_\mathrm{b}$.}
	\label{fig:experimental_setup}
\end{figure}

The~experimental setup employed in the~present work is shown in \cref{fig:experimental_setup}. An~Yb:YAG disk laser (Trumpf TruDisk 10001) that was connected to a~fibre with a~core diameter of $\SI{6e-4}{\meter}$ was employed. The~fibre transports the~laser light towards the~focusing optics (Trumpf BEO D70), consisting of a~$\SI{200}{\milli\meter}$ collimator and $\SI{400}{\milli\meter}$ focusing lens. The~focusing optics are mounted to a~6-DOF robot (ABB IRB-2600M2004). Using this setup, a~laser spot with a~diameter of $\SI{1.2}{\milli\meter}$ and a~top-hat power-density distribution was produced (see \cref{fig:experimental_setup}(b)). Melting tracks with a~length of $\SI{80}{\milli\meter}$ were made on an~AISI~316L plate with dimensions of $\SI{250}{\milli\meter} \times \SI{100}{\milli\meter} \times \SI{10}{\milli\meter}$. The~travel speed was set to $\SI{20}{\milli\meter\per\second}$. {The~melt pool was protected from oxidation during the~process using argon gas at a~flow rate of about $\SI{20}{\liter\per\min}$.} Each experiment was repeated at least three times to ensure that the~results are reproducible. The~samples were cut transversely, polished and etched to capture macrographs using a~digital microscope (Keyence VHX 7000). {A~solution of $\SI{100}{\milli\liter}$ \ch{HCl}, $\SI{100}{\milli\liter}$ \ch{H2O} and $\SI{10}{\milli\liter}$ \ch{NHO3}} with a~temperature of about $\SI{310}{\kelvin}$ was used for etching the~samples.

\FloatBarrier

\section{Results and discussion}
\label{sec:results}

\subsection{Model validation}
\label{sec:validation}
\begin{figure}[!htb] 
	\centering
	\includegraphics[width=1.0\linewidth]{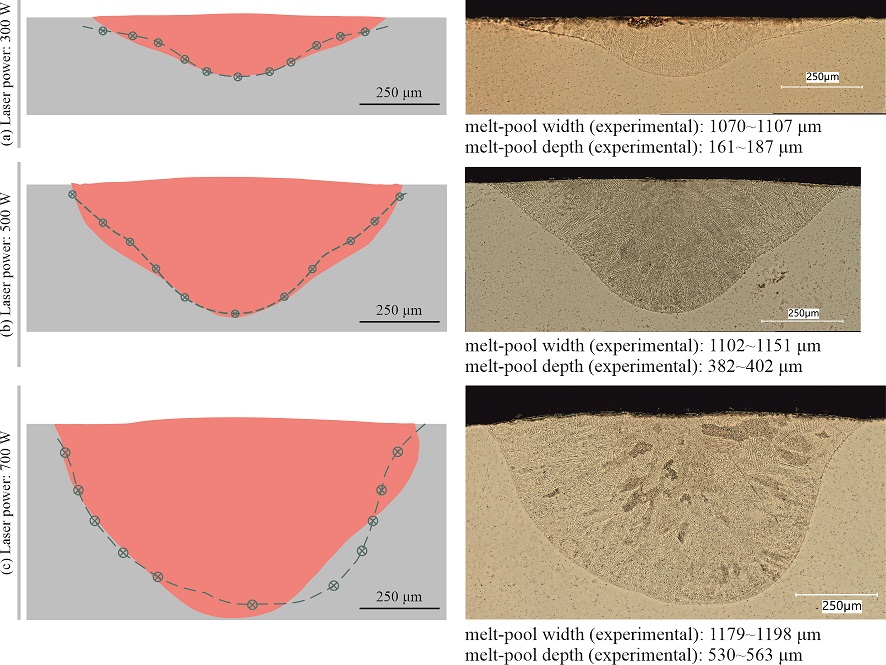}
	\caption{{Numerically predicted melt-pool shapes obtained from the~present computational model (left column, regions shaded in pink) compared with the~experimental macrographs (right column, circles and dashed lines). The~base material is stainless steel 316L. An~Yb:YAG laser ($\lambda = \SI{1.030e-6}{\meter}$) was used, the~laser beam had a~top-hat power-density distribution, the~spot size $d_\mathrm{b}$ was $\SI{1.2}{\milli\meter}$ and travel speed $\mathscr{V}$ was set to $\SI{20}{\milli\meter\per\second}$. The~energy-density ($\mathscr{E} = \mathscr{P}/(\mathscr{V}d_\mathrm{b})$) ranges between $\SI{12.5}{\mega\joule\per\square\meter}$ and $\SI{29.2}{\mega\joule\per\square\meter}$.}}
	\label{fig:fibre_laser_melting_valid}
\end{figure}

The~reliability and accuracy of the~present computational model are examined by comparing the~numerically predicted melt-pool shapes with those obtained from experiments with different laser systems, laser powers and power-density distributions. To visualise the~melt-pool shapes, cross-sectional macrographs were prepared \textit{ex situ} after experiments and iso-surfaces of solidus temperature were projected on the~$x$-$z$ plane after numerical simulations. It should be noted that the~numerical results were obtained using the~variable absorptivity model described in~\cref{sec:absorptivity_model} without calibration. \Cref{fig:fibre_laser_melting_valid} shows a~comparison between the~melt-pool shapes obtained from the~present computational model with those obtained from experiments using an~Yb:YAG laser ($\lambda = \SI{1.030e-6}{\meter}$) and different laser powers, which indicates a~reasonable agreement (generally less than~5\% difference in melt-pool dimensions).

\begin{figure}[!htb] 
	\centering
	\includegraphics[width=1.0\linewidth]{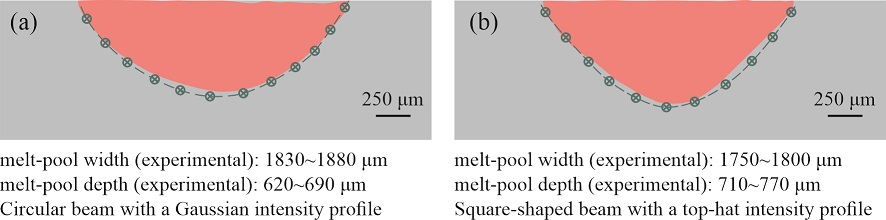}
	\caption{{Comparison of the~melt-pool shapes obtained from the~present computational model (regions shaded in pink) with the~experimental measurements of~Kell~\textit{et al.}~\cite{Kell_2006} (circles and dashed line). (a)~Circular laser beam with a~Gaussian power-density distribution and a~spot size (D4$\sigma$) of $d_\mathrm{b} = \SI{1.25}{\milli\meter}$, and (b) Square-shaped laser beam with a~top-hat power-density distribution and a~spot size of $d_\mathrm{b} = \SI{1.25}{\milli\meter}$. The~base material is stainless steel 316L. A~\ch{CO2} laser ($\lambda = \SI{1.060e-5}{\meter}$) was used and the~energy-density ($\mathscr{E} = \mathscr{P}/(\mathscr{V}d_\mathrm{b})$) was set to $\SI{120}{\mega\joule\per\square\meter}$ for both cases.}}
	\label{fig:co2_laser_melting_valid}
\end{figure}

The~characteristics of the~laser system used in laser melting can affect the~absorptivity and hence can change the~resulting melt-pool shape. The~results of the~present computational model are also benchmarked against the~experimental data reported by~Kell~\textit{et al.}~\cite{Kell_2006} for laser melting of a~$\SI{1}{\milli\meter}$-thick steel plate using a~\ch{CO2} laser ($\lambda = \SI{1.060e-5}{\meter}$) with the~energy-density ($\mathscr{E} = \mathscr{P}/(\mathscr{V}d_\mathrm{b})$) being set to $\SI{120}{\mega\joule\per\square\meter}$, and the~results are shown in~\cref{fig:co2_laser_melting_valid}. {To compare the~numerically predicted melt-pool shape with experimental measurements, the~relative difference between melt-pool dimensions (\textit{i.e.}~the~melt-pool width and depth) was calculated as follows:}

\begin{equation}
	\%\mathrm{Deviation} = \left|\frac{\mathscr{L}_{\mathrm{numerical}} - \mathscr{L}_{\mathrm{experimental}}}{\mathscr{L}_{\mathrm{experimental}}}\right| \times 100,
	\label{eq:deviation}
\end{equation}

\noindent
{where, $\mathscr{L}$ indicates the~melt-pool depth and width.}
In this case, the~deviation between the~numerically predicted and the~experimentally measured melt-pool dimensions is less than 2\%, demonstrating the~reliability of the~present computational model in predicting the~melt-pool shape. The~deviation between the~numerical and experimental results can be attributed to uncertainties in modelling temperature-dependent material properties, particularly in the~liquid phase, the~assumptions made to develop the~present computational model as well as uncertainties associated with the~experimental measurements.	

\FloatBarrier

\subsection{Melt-pool shape and dimensions}
\label{sec:shape}
\begin{figure}[!htb] 
	\centering
	\includegraphics[width=1.0\linewidth]{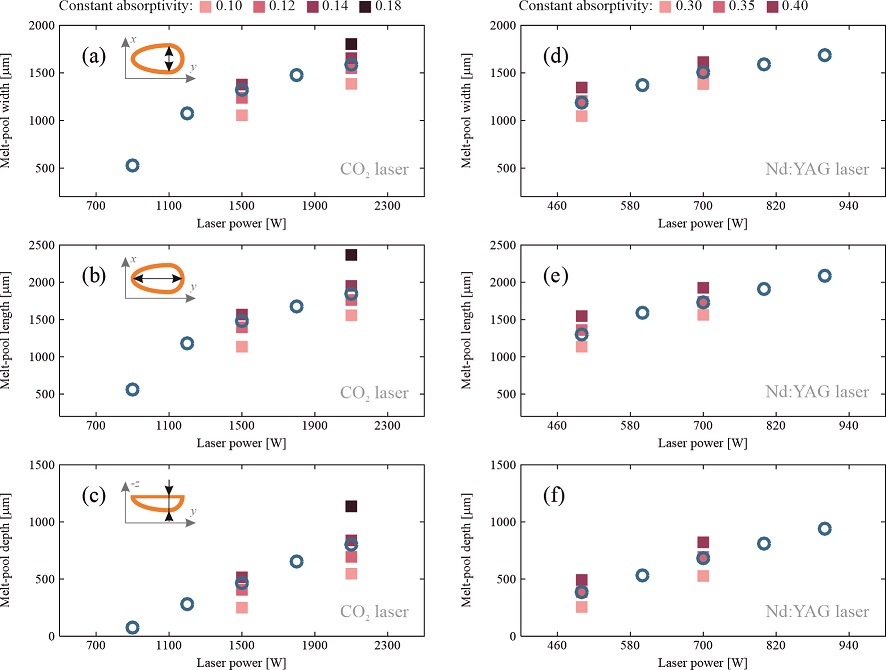} 
	\caption{Melt-pool dimensions obtained from the~present computational model using variable absorptivity (circles) and constant absorptivity (squares) for different laser powers and laser systems ((a--c)~\ch{CO2} laser \mbox{($\lambda = \SI{1.060e-5}{\meter}$)} and (d--f)~Nd:YAG laser ($\lambda = \SI{1.064e-6}{\meter}$)). Travel speed $\mathscr{V}$ was set to $\SI[parse-numbers=false]{10^{-2}}{\meter\per\second}$ for all the~cases.}
	\label{fig:meltpool_dimensions}
\end{figure}

{To be able to systematically study the~effects of laser characteristics and melt-pool surface deformations on variation of local absorptivity, three batches of simulations are considered for different laser types. For cases in batch 1 and 2, the~power density is too low to cause significant vaporisation and surface deformations are small compared to the~melt-pool depth. Thus, changes in the~absorptivity for a~specific laser and material can be attributed primarily to changes in surface temperature. The~laser spot size for the~cases in batch 3 is intentionally chosen smaller than that for the~cases in batch 1 and 2 to achieve high values of power-density, resulting in significant vaporisation of the~material and melt-pool surface deformations compared to its depth. For all three batches, the~results obtained using the enhanced absorption model are compared with those obtained using a~constant absorptivity.}
\Cref{fig:meltpool_dimensions} shows the~numerically predicted melt-pool dimensions obtained for different laser powers using \ch{CO2} and Nd:YAG lasers (\textit{i.e.}~cases in batch 1 and 2). The~melt-pool dimensions obtained using the~variable absorptivity model are compared with those obtained using different constant values of the~absorptivity. For cases in batch 1 and 2, the~power density is too low to cause significant vaporisation and surface deformations are small compared to the~melt-pool depth. Thus, changes in the~absorptivity for a~specific laser and material can be attributed primarily to changes in surface temperature (see \cref{fig:absorptivity}).

For the~cases where the~\ch{CO2} laser was employed (\cref{fig:meltpool_dimensions}(a--c)), melt-pool dimensions predicted using a~constant absorptivity between 0.12 and 0.14 seem to agree with those obtained using the~variable absorptivity model. However, the~results suggest that employing a~constant absorptivity does not necessarily render all the~melt-pool dimensions with the~same level of accuracy, which means the~results are less reliable with respect to those obtained using the~variable absorptivity model. This can be attributed to the~fact that changes in local energy absorption due to changes in surface temperature, and changes in total energy absorption over time are both neglected when a~constant absorptivity is employed. Surface temperature in the~spot region after reaching a~quasi-steady-state condition for \ch{CO2} laser melting with a~laser power of $\mathscr{P} = \SI{2100}{\watt}$ ranges between $\SI{1900}{\kelvin}$ and $\SI{2650}{\kelvin}$, resulting in absorptivities that range between 0.130 and 0.136 according to the~variable absorptivity model and in agreement with the~results shown in \cref{fig:meltpool_dimensions}(a--c). Although a~good agreement between numerical and experimental melt-pool dimensions might be achievable using a~constant absorptivity model, the~use of a~constant absorptivity requires a~\textit{posteriori} fitting of the~absorptivity value to the~experiments, whereas such a~fitting is not required employing the~variable absorptivity model described in~\cref{sec:absorptivity_model}. 

The~results shown in \cref{fig:meltpool_dimensions} suggest that for a~certain set of process parameters, a~lower laser power is required to obtain a~melt-pool with similar dimensions using an~Nd:YAG laser with an~emission wavelength of $\lambda = \SI{1.064e-6}{\meter}$ than a~\ch{CO2} laser with an~emission wavelength of $\lambda = \SI{1.060e-5}{\meter}$. This arises because the~absorptivity for a~\ch{CO2} laser is generally lower than that for an~Nd:YAG laser when the~incident angle is too small to affect the~absorptivity significantly ($\theta < 40^\circ$, as is suggested by the~data shown in \cref{fig:absorptivity}), which is the~case in conduction-mode laser melting. For the~cases where the~Nd:YAG laser was employed (\cref{fig:meltpool_dimensions}(d--f)), using a~constant absorptivity of 0.35 can render the~melt-pool dimensions with a~reasonable resolution. When an~Nd:YAG laser with a~laser power of $\mathscr{P} = \SI{700}{\watt}$ is employed, numerically predicted surface temperature in the~spot region after reaching a~quasi-steady-state condition ranges between $\SI{1900}{\kelvin}$ and $\SI{2600}{\kelvin}$. For this temperature range, the~absorptivity varies between 0.347 and 0.36 according to the~variable absorptivity model, and its arithmetic average 0.354 is close to 0.35. Since the~melt-pool surface temperature and its distribution are not known \textit{a~priory} and are significantly influenced by the~process parameters as well as the~complex internal molten metal flow, running trial-and-error tests is indispensable to calibrate the~value of constant absorptivity. Running such trial-and-error tests increases the~total costs of computational analyses and such \textit{ad hoc} calibration often lacks generality.

\begin{figure}[!htb] 
	\centering
	\includegraphics[width=1.0\linewidth]{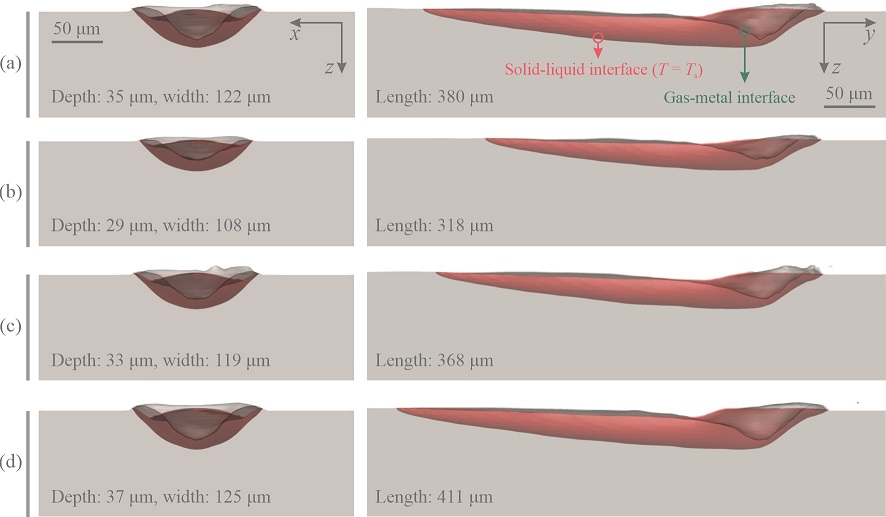}
	\caption{Melt-pool shapes obtained from the~present computational model for different cases in batch 3 (fibre laser ($\lambda = \SI{1.070e-6}{\meter}$), $\mathscr{P} = \SI{200}{\watt}$, spot size $d_\mathrm{b} = \SI{1.1e-4}{\meter}$ and travel speed $\mathscr{V} = \SI{1.5}{\meter\per\second}$). (a)~variable absorptivity, (b)~constant absorptivity~$\mathscr{a} = 0.30$, (c)~$\mathscr{a} = 0.35$ and (d)~$\mathscr{a} = 0.40$.}
	\label{fig:meltpool_shape_fibre}
\end{figure}

\Cref{fig:meltpool_shape_fibre} shows the~numerically predicted melt-pool shapes obtained using both variable and constant absorptivity models for a~fibre laser with an~emission wavelength of $\lambda = \SI{1.070e-6}{\meter}$ (\textit{i.e.}~cases in batch 3). The~power density for the~cases in batch 3 is relatively high, resulting in significant vaporisation of the~material and melt-pool surface deformations compared to its depth. In~contrast to the~cases in batch 1 and 2, the~absorptivity for the~cases in batch 3 are affected by both temperature and incident angle of the~laser ray (see \cref{fig:absorptivity}).

The~results presented in \cref{fig:meltpool_shape_fibre} show an agreement between the~melt-pool dimensions predicted using a~constant absorptivity of 0.35 and those obtained using the~variable absorptivity model. For the~cases in batch 3, surface temperature in the~laser spot region ranges roughly from $\SI{800}{\kelvin}$ (because of the~relatively high travel speed, the~material is in the~solid state in front part of the~laser spot region) to the~boiling temperature of $\SI{3086}{\kelvin}$  corresponding to absorptivities between 0.30 and 0.37 (with an~arithmetic average of 0.335) based on the~variable absorptivity model for $\theta = 0^\circ$. Demonstrably, reducing the~absorptivity from 0.35 to 0.335 decreases the~total amount of energy absorbed by the~material, resulting in smaller melt-pool dimensions than those predicted using the~variable absorptivity model. The~increase in local energy absorption due to the~increase in temperature and laser incident angle is neglected when a~constant absorptivity model is employed. For the~cases in batch 3, the~incident angle of the~laser ray $\theta$ increases from $0^\circ$ to $50^\circ$ with melt-pool surface depression, resulting in an~increase in the~local energy absorption according to the~variable absorptivity model and in turn increases the~melt-pool surface temperature, which leads to further increase in absorptivity. Eventually, the~material reaches the~boiling temperature and vaporisation limits further increase of melt-pool surface temperature. Variations of total energy input and energy-density distribution due to dynamic changes of surface temperature and morphology affect material vaporisation and thus the~recoil pressure that is responsible for melt-pool surface depression. Consequently, these effects cannot be described adequately when a~constant absorptivity model is employed in numerical simulations of laser welding and additive manufacturing. Modelling such phenomena with sufficient accuracy is crucial in numerical simulations of transition from conduction to keyhole mode laser melting as well as those developed to predict solidification microstructure and texture.

\FloatBarrier

\subsection{Thermal and fluid flow fields}
\label{sec:thermal_fluid}
\begin{figure}[!htb] 
	\centering
	\includegraphics[width=0.85\linewidth]{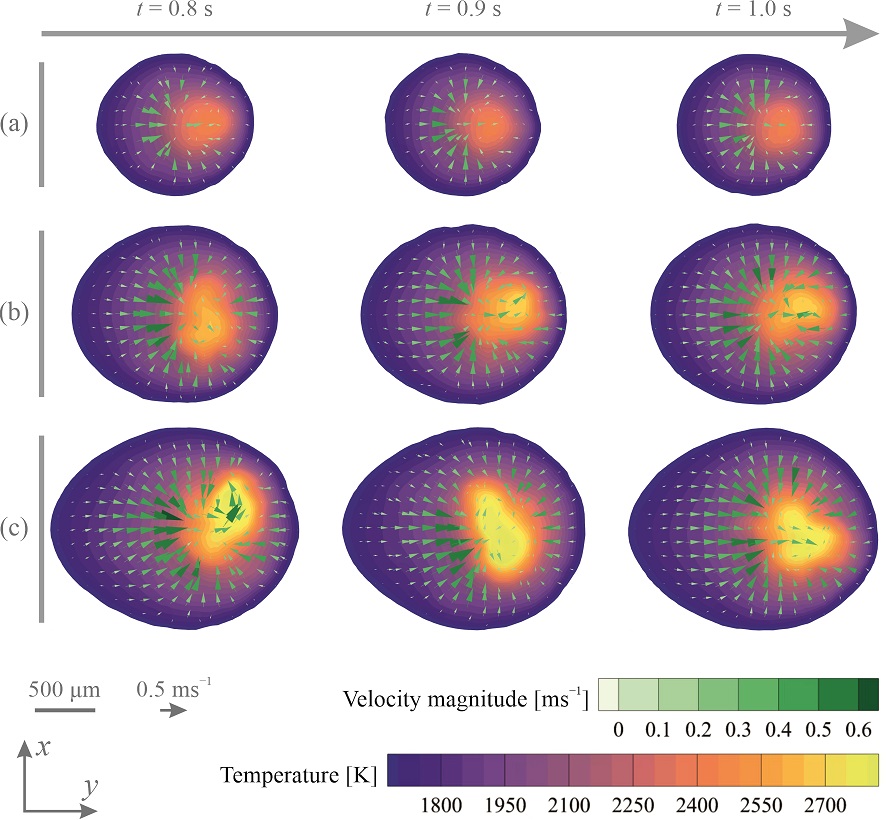}
	\caption{Evolution of thermal and fluid flow fields over the~melt-pool surface during laser melting of stainless steel 316L with different laser powers. (a) $\mathscr{P} = \SI{500}{\watt}$, (b) $\mathscr{P} = \SI{700}{\watt}$, and (c) $\mathscr{P} = \SI{900}{\watt}$. The~variable absorptivity model is utilised. Cases belong to batch 2, where an~Nd:YAG laser ($\lambda = \SI{1.064e-6}{\meter}$) is employed with a~travel speed of $\SI[parse-numbers=false]{10^{-2}}{\meter\per\second}$.}
	\label{fig:thermal_fluid_ndyag}
\end{figure}

Soon after exposing the~material to laser radiation, a~melt pool forms and grows over time and if the~boundary conditions allow, reaches a~quasi-steady-state condition. \Cref{fig:thermal_fluid_ndyag} shows the~thermal and fluid flow fields over the~melt-pool surface at different time instances after reaching the~quasi-steady-state condition during laser melting using an~Nd:YAG laser ($\lambda = \SI{1.064e-6}{\meter}$) with different laser powers. The~temperature gradient induced over the~surface generates Marangoni shear forces that drive molten metal flow. The~molten metal moves from the~cold regions adjacent to the~melt-pool rim towards the~central part of the~pool while absorbing energy from the~laser beam. This agrees well with experimental observations and discussions reported by Mills~\textit{et al.}~\cite{Mills_1998} for stainless steel alloys. The~absorbed energy advects with the~flow and diffuses through the~material into the~surrounding solid regions. As~the~material absorbs energy, surface temperature increases and if the~power-density is high enough, the~surface temperature reaches a~critical value at which the~sign of the~temperature gradient of surface tension ($\mathrm{d} \gamma / \mathrm{d} T$) changes (see \cref{fig:materials_properties}(e)), resulting in a~change in flow direction. Interactions between the~inward and the~outward streams result in a~complex flow pattern in the~pool that is inherently unsteady and three-dimensional~\cite{Kidess_2016_PhysFlu,Ebrahimi_2020}. Two vortices are observed over the~melt-pool surface close to the~hot spot, generating an~asymmetric flow pattern that fluctuates around the~centre-line of the~pool. This fluid motion forms a~rotational flow pattern in the~pool that transfers the~absorbed heat from the~surface to the~bottom of the~pool~\cite{Zhao_2010}. Because of this rotational fluid motion, an~element of molten metal volume may move from one side of the~pool to the~other side, resulting in a~cross-cellular flow~\cite{Schatz_2001} that enhances mixing in the~pool. Maximum fluid velocity in the~pool reach values of about $\SI{0.6}{\meter\per\second}$, corresponding to a~P\'eclet number ($\mathrm{Pe} = \rho c_\mathrm{p} \mathscr{D} \lVert \mathbf{u} \rVert / k$) in the~order of $\mathscr{O}(100)$ that indicates the~significant contribution of advection to the~total energy transfer. Molten metal flow in the~pool disturbs the~thermal field and in turn affects the~absorptivity and Marangoni forces. A~similar flow pattern is observed over the~surface when a~\ch{CO2} laser is employed, and representative results are provided in the~supplementary materials.

\begin{figure}[!htb] 
	\centering
	\includegraphics[width=1.0\linewidth]{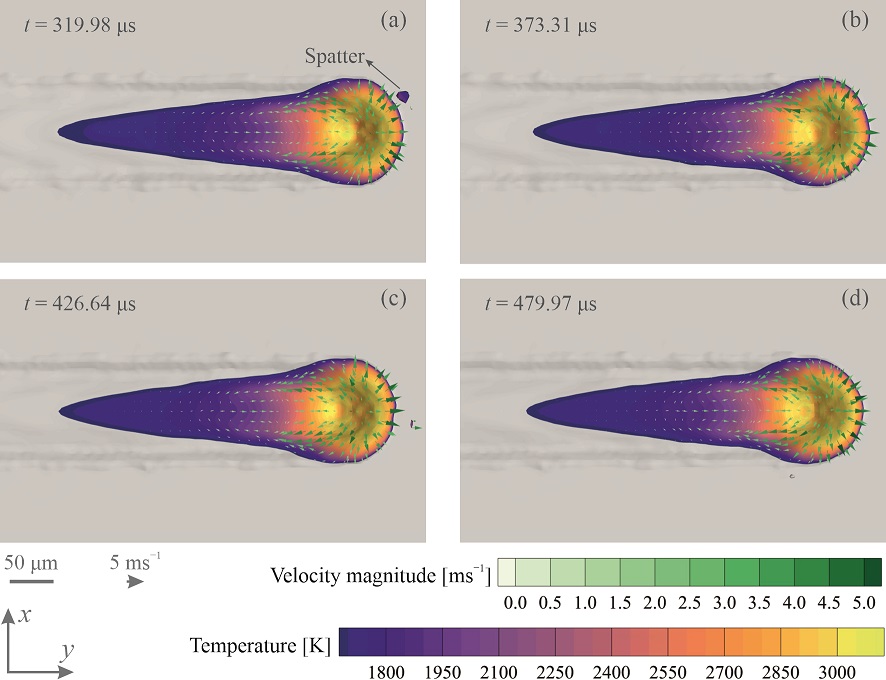}
	\caption{Evolution of thermal and fluid flow fields over the~melt-pool surface during laser melting of stainless steel 316L using a~fibre laser ($\lambda = \SI{1.070e-6}{\meter}$) with a~laser power of $\mathscr{P} = \SI{200}{\watt}$ and a~travel speed of $ \SI{1.5}{\meter\per\second}$. The~case belong to batch 3 and the~variable absorptivity model is utilised.}
	\label{fig:thermal_fluid_fibre}
\end{figure}

\Cref{fig:thermal_fluid_fibre} shows a~time series of numerically predicted thermal and fluid flow fields over the~melt-pool surface after reaching the~quasi-steady-state condition during laser melting using a~fibre laser ($\lambda = \SI{1.070e-6}{\meter}$) with a~laser power of $\mathscr{P} = \SI{200}{\watt}$ and a~travel speed of $\mathscr{V} = \SI{1.5}{\meter\per\second}$. In~this case, three distinctive regions are identified: a~region with significant surface depression under the~effect of recoil pressure, a~trailing region characterised by low fluid velocities (less than $\SI{0.2}{\meter\per\second}$) and temperature (less than $\SI{1750}{\kelvin}$), and a~transition zone in between. A~similar choice of subdivision is reported by Khairallah~\textit{et al.}~\cite{Khairallah_2016} to describe the~anatomy of a~melt track in selective laser melting of a~powder bed, where the~surface tension temperature gradient ($\mathrm{d} \gamma / \mathrm{d} T$) was assumed to be a~negative constant value. Because of the~relatively high laser power-density, surface temperature in the~spot region increases rapidly and reaches the~boiling temperature $T_\mathrm{b}$, leading to significant material vaporisation and increase in recoil pressure that locally deforms the~melt-pool surface. Beneath the~front part of the~depressed region, a~relatively thin layer of molten metal exists as shown in \cref{fig:meltpool_shape_fibre}. The~maximum molten metal velocity over the~surface in the~depressed region is about $\SI{6}{\meter\per\second}$ due to the~large temperature gradients, forming a~multi-cellular flow pattern in the~thin molten metal layer due to Marangoni flow instabilities~\cite{Schatz_2001}. {The~maximum molten metal velocity predicted for cases in batch 3 is higher than that for cases in batch 1 and 2. This is primarily attributed to larger temperature gradients induced over the~surface, increasing the~magnitude of Marangoni shear force. Moreover, for temperatures above a~critical value at which the~sign of the~temperature gradient of surface tension ($\mathrm{d} \gamma / \mathrm{d} T$) changes from positive to negative (see \cref{fig:materials_properties}(e)), the~absolute value of the temperature gradient of surface tension increases with temperature, increasing the magnitude of Marangoni force applied to the molten material.} Due to the~recoil pressure and the~outward fluid motion on the~surface, molten metal accumulates ahead of the~depressed region, which is also observed experimentally by~Nakamura~\textit{et al.}~\cite{Nakamura_2015} and simulated numerically by~\mbox{Khairallah~\textit{et al.}~\cite{Khairallah_2016,Khairallah_2020}}. Elements of the~accumulated liquid volume can be ejected from the~pool and form spatters as shown in \cref{fig:thermal_fluid_fibre}(a~and~c). Spatters are small compared to the~melt-pool volume and generally cool down during their flight and thus do not have sufficient thermal energy to melt the~substrate and stick to the~surface. Molten metal moves from the~central region of the~depressed region towards the~melt-pool rim and transfers the~heat absorbed from the~laser. This fluid velocity corresponds to a~P\'eclet~number~($\mathrm{Pe}$) in the~order of $\mathscr{O}(100)$, which is similar to the~conduction-mode laser melting. Molten metal moving from the~depressed region towards the~rear part of the~pool meets an~inward flow in the~transition region, resulting in the~formation of two vortices over the~surface. In the~transition region, surface temperature is less than the~critical temperature at which the~sign of the~temperature gradient changes (see \cref{fig:materials_properties}(e)), thus the~surface tension increases with increasing the~temperature (\textit{i.e.}~$\mathrm{d} \gamma / \mathrm{d} T > 0$) in the~transition region, resulting in a~fluid motion from the~cold to the~hot regions. In the~trailing region, temperature gradients are too small to generate significant Marangoni forces to drive the~molten metal flow, and thus thermal diffusion dominates the~energy transfer.

\FloatBarrier

\section{Conclusions}
\label{sec:conclusions}
The~influence of laser characteristics on internal molten metal flow in laser-beam melting of a~metallic substrate was investigated numerically using a~high-fidelity three-dimensional model. An~enhanced absorption model that accounts for laser emission wavelength, surface temperature, laser-ray incident angle and material composition was utilised in the~model, and the~results compared with experimental measurements as well as numerical data predicted using a~constant absorption model. The~physics of complex heat and molten metal flow in laser melting is described for various test cases with different laser powers, laser emission wavelengths, and power-density distributions.

For conduction-mode laser melting, where surface deformations are small compared to the~melt-pool depth, the~absorptivity changes primarily because of changes in surface temperature. However, for cases that surface deformations are significant with respect to the~melt-pool depth, changes in the~absorptivity are affected by both the~surface temperature and the~laser-ray incident angle. Changes in the~absorptivity affect energy-density distribution over the~surface and hence the~thermal field over the~melt-pool surface, which in turn can influence the~Marangoni-driven molten metal flow as well as the~distribution of recoil pressure over the~surface. These physical processes are tightly coupled to one another, resulting in highly non-linear responses to changes in process parameters. 

{For laser melting processes with a~relatively low power density using a~\ch{CO2} or fibre transmissible laser (with an~emission wavelength close to \SI{1}{\micro\meter}), the~molten metal velocities and surface deformations are relatively small. Because of the~small incident angle, the~absorptivity for a~\ch{CO2} laser is lower than that for a~Nd:YAG laser; thus, a~lower laser power is required to obtain a~melt-pool with similar dimensions using an~Nd:YAG laser as compared to a~\ch{CO2} laser. Switching to a~relatively high power density laser melting process, molten metal velocities increase compared to the~low power density processes. For sufficiently high power densities, melt-pool surface deformations become significant, resulting in strongly enhanced laser absorption which in turn further enhances metal vaporisation.}

The~results of the~present work demonstrate that the~coupling between these physical processes cannot be rendered with sufficient resolution employing a~constant absorptivity model, reducing the~range of predictability of the~computational models developed to describe the~dynamics of melt-pool behaviour in laser welding and additive manufacturing. Moreover, considering absorptivity as a~calibration parameter in computational models necessitates trial-and-error simulations, which increases the~total costs of computational analyses.

{Although the~focus of the~present work is primarily on laser melting of bare metallic substrates without powder layers, the~fundamental laser-matter interaction mechanisms described here are similar to those in laser melting of powder beds. The~enhanced laser-absorptivity model employed in the~present work can also be utilised in numerical simulations of melt-pool behaviour in laser melting of powder beds, provided that multiple reflections are included in the~model. }

\section*{Acknowledgement}
\label{sec:acknowledgement}

This research was carried out under project numbers F31.7.13504, P16-46/S17024i and P16-46/S17024m in the~framework of the~Partnership Program of the~Materials innovation institute M2i (www.m2i.nl) and the~Foundation for Fundamental Research on Matter (FOM) (www.fom.nl), which is part of the~Netherlands Organisation for Scientific Research (www.nwo.nl). This research project is also a~part of Aim2XL program (www.m2i.nl/aim2xl). The~authors would like to thank the~industrial partners in this project ``Allseas Engineering B.V." and ``Rotterdam Fieldlab Additive Manufacturing B.V. (RAMLAB)" for the~financial support. 

\section*{Author Contributions}
\label{sec:author_contributions}

Conceptualisation, A.E.; methodology, A.E., M.S., S.J.L.B. and M.L.; software, A.E.; validation, A.E., M.S. and S.J.L.B.; formal analysis, A.E.; investigation, A.E., M.S., S.J.L.B. and M.L.; resources, A.E., M.S., S.J.L.B., M.L., G.R.B.E.R. and M.J.M.H.; data curation, A.E., M.S., S.J.L.B. and M.L.; writing---original draft preparation, A.E.; writing---review and editing, A.E., M.S., S.J.L.B., M.L., G.R.B.E.R., I.M.R., C.R.K. and M.J.M.H.; visualisation, A.E., M.S., S.J.L.B. and M.L.; supervision, G.R.B.E.R., I.M.R., C.R.K. and M.J.M.H.; project administration, A.E., G.R.B.E.R. and M.J.M.H.; and funding acquisition, G.R.B.E.R., I.M.R. and M.J.M.H.

{
	\small
	\onehalfspacing

}
\bibliographystyle{article-bibstyle} 

\begin{thebibliography}{62}
		\providecommand{\natexlab}[1]{#1}
		\providecommand{\url}[1]{\texttt{#1}}
		\expandafter\ifx\csname urlstyle\endcsname\relax
		\providecommand{\doi}[1]{doi: #1}\else
		\providecommand{\doi}{doi: \begingroup \urlstyle{rm}\Url}\fi
		
		\bibitem[Khairallah \textit{et~al.}(2020)Khairallah, Martin, Lee, Guss, Calta,
		Hammons, Nielsen, Chaput, Schwalbach, Shah, Chapman, Willey, Rubenchik,
		Anderson, Wang, Matthews, and King]{Khairallah_2020}
		Khairallah, S.~A., Martin, A.~A., Lee, J. R.~I., Guss, G., Calta, N.~P.,
		Hammons, J.~A., Nielsen, M.~H., Chaput, K., Schwalbach, E., Shah, M.~N.,
		Chapman, M.~G., Willey, T.~M., Rubenchik, A.~M., Anderson, A.~T., Wang,
		Y.~M., Matthews, M.~J., and King, W.~E.
		\newblock Controlling interdependent meso-nanosecond dynamics and defect
		generation in metal {3D} printing.
		\newblock \emph{Science}, 368\penalty0 (6491):\penalty0 660--665, 2020.
		\newblock \doi{10.1126/science.aay7830}.
		
		\bibitem[Markl and Körner(2016)]{Markl_2016}
		Markl, M. and Körner, C.
		\newblock Multiscale modeling of powder bed{\textendash}based additive
		manufacturing.
		\newblock \emph{Annual Review of Materials Research}, 46\penalty0 (1):\penalty0
		93--123, 2016.
		\newblock \doi{10.1146/annurev-matsci-070115-032158}.
		
		\bibitem[Ebrahimi \textit{et~al.}(2021{\natexlab{a}})Ebrahimi, Kleijn, and
		Richardson]{Ebrahimi_2021}
		Ebrahimi, A., Kleijn, C.~R., and Richardson, I.~M.
		\newblock A simulation-based approach to characterise melt-pool oscillations
		during gas tungsten arc welding.
		\newblock \emph{International Journal of Heat and Mass Transfer}, 164:\penalty0
		120535, 2021{\natexlab{a}}.
		\newblock \doi{10.1016/j.ijheatmasstransfer.2020.120535}.
		
		\bibitem[Francois \textit{et~al.}(2017)Francois, Sun, King, Henson, Tourret,
		Bronkhorst, Carlson, Newman, Haut, Bakosi, Gibbs, Livescu, Wiel, Clarke,
		Schraad, Blacker, Lim, Rodgers, Owen, Abdeljawad, Madison, Anderson,
		Fattebert, Ferencz, Hodge, Khairallah, and Walton]{Francois_2017}
		Francois, M.~M., Sun, A., King, W.~E., Henson, N.~J., Tourret, D., Bronkhorst,
		C.~A., Carlson, N.~N., Newman, C.~K., Haut, T., Bakosi, J., Gibbs, J.~W.,
		Livescu, V., Wiel, S. A.~V., Clarke, A.~J., Schraad, M.~W., Blacker, T., Lim,
		H., Rodgers, T., Owen, S., Abdeljawad, F., Madison, J., Anderson, A.~T.,
		Fattebert, J.-L., Ferencz, R.~M., Hodge, N.~E., Khairallah, S.~A., and
		Walton, O.
		\newblock Modeling of additive manufacturing processes for metals: {Challenges}
		and opportunities.
		\newblock \emph{Current Opinion in Solid State and Materials Science},
		21\penalty0 (4):\penalty0 198--206, 2017.
		\newblock \doi{10.1016/j.cossms.2016.12.001}.
		
		\bibitem[DebRoy \textit{et~al.}(2020)DebRoy, Mukherjee, Wei, Elmer, and
		Milewski]{DebRoy_2020}
		DebRoy, T., Mukherjee, T., Wei, H.~L., Elmer, J.~W., and Milewski, J.~O.
		\newblock Metallurgy, mechanistic models and machine learning in metal
		printing.
		\newblock \emph{Nature Reviews Materials}, 6\penalty0 (1):\penalty0 48--68,
		2020.
		\newblock \doi{10.1038/s41578-020-00236-1}.
		
		\bibitem[Cook and Murphy(2020)]{Cook_2020}
		Cook, P.~S. and Murphy, A.~B.
		\newblock Simulation of melt pool behaviour during additive manufacturing:
		Underlying physics and progress.
		\newblock \emph{Additive Manufacturing}, 31:\penalty0 100909, 2020.
		\newblock \doi{10.1016/j.addma.2019.100909}.
		
		\bibitem[Simonds \textit{et~al.}(2021)Simonds, Tanner, Artusio-Glimpse,
		Williams, Parab, Zhao, and Sun]{Simonds_2021}
		Simonds, B.~J., Tanner, J., Artusio-Glimpse, A., Williams, P.~A., Parab, N.,
		Zhao, C., and Sun, T.
		\newblock The causal relationship between melt pool geometry and energy
		absorption measured in real time during laser-based manufacturing.
		\newblock \emph{Applied Materials Today}, 23:\penalty0 101049, 2021.
		\newblock \doi{10.1016/j.apmt.2021.101049}.
		
		\bibitem[Ebrahimi \textit{et~al.}(2021{\natexlab{b}})Ebrahimi, Kleijn, and
		Richardson]{Ebrahimi_2020}
		Ebrahimi, A., Kleijn, C.~R., and Richardson, I.~M.
		\newblock Numerical study of molten metal melt pool behaviour during
		conduction-mode laser spot melting.
		\newblock \emph{Journal of Physics D: Applied Physics}, 54:\penalty0 105304,
		2021{\natexlab{b}}.
		\newblock \doi{10.1088/1361-6463/abca62}.
		
		\bibitem[Ebrahimi \textit{et~al.}(2021{\natexlab{c}})Ebrahimi, Kleijn, Hermans,
		and Richardson]{Ebrahimi_2021_b}
		Ebrahimi, A., Kleijn, C.~R., Hermans, M. J.~M., and Richardson, I.~M.
		\newblock The effects of process parameters on melt-pool oscillatory behaviour
		in gas tungsten arc welding.
		\newblock \emph{Journal of Physics D: Applied Physics}, 54\penalty0
		(27):\penalty0 275303, 2021{\natexlab{c}}.
		\newblock \doi{10.1088/1361-6463/abf808}.
		
		\bibitem[Xie \textit{et~al.}(1997)Xie, Kar, Rothenflue, and Latham]{Xie_1997}
		Xie, J., Kar, A., Rothenflue, J.~A., and Latham, W.~P.
		\newblock Temperature-dependent absorptivity and cutting capability of
		{CO}{\textsubscript{2}}, {Nd}:{YAG} and chemical oxygen{\textendash}iodine
		lasers.
		\newblock \emph{Journal of Laser Applications}, 9\penalty0 (2):\penalty0
		77--85, 1997.
		\newblock \doi{10.2351/1.4745447}.
		
		\bibitem[Mahrle and Beyer(2009)]{Mahrle_2009}
		Mahrle, A. and Beyer, E.
		\newblock Theoretical aspects of fibre laser cutting.
		\newblock \emph{Journal of Physics D: Applied Physics}, 42\penalty0
		(17):\penalty0 175507, 2009.
		\newblock \doi{10.1088/0022-3727/42/17/175507}.
		
		\bibitem[Ren \textit{et~al.}(2021)Ren, Zhang, Fu, Jiang, and Zhao]{Ren_2021}
		Ren, Z., Zhang, D.~Z., Fu, G., Jiang, J., and Zhao, M.
		\newblock High-fidelity modelling of selective laser melting copper alloy:
		Laser reflection behavior and thermal-fluid dynamics.
		\newblock \emph{Materials {\&} Design}, 207:\penalty0 109857, 2021.
		\newblock \doi{10.1016/j.matdes.2021.109857}.
		
		\bibitem[Yang \textit{et~al.}(2021)Yang, Bauerei{\ss}, Markl, and
		Körner]{Yang_2021}
		Yang, Z., Bauerei{\ss}, A., Markl, M., and Körner, C.
		\newblock Modeling laser beam absorption of metal alloys at high temperatures
		for selective laser melting.
		\newblock \emph{Advanced Engineering Materials}, page 2100137, 2021.
		\newblock \doi{10.1002/adem.202100137}.
		
		\bibitem[Shu \textit{et~al.}(2021)Shu, Galles, Tertuliano, McWilliams, Yang,
		Cai, and Lew]{Shu_2021}
		Shu, Y., Galles, D., Tertuliano, O.~A., McWilliams, B.~A., Yang, N., Cai, W.,
		and Lew, A.~J.
		\newblock A critical look at the prediction of the temperature field around a
		laser-induced melt pool on metallic substrates.
		\newblock \emph{Scientific Reports}, 11\penalty0 (1), 2021.
		\newblock \doi{10.1038/s41598-021-91039-z}.
		
		\bibitem[Kidess \textit{et~al.}(2016{\natexlab{a}})Kidess, Kenjere{\v{s}},
		Righolt, and Kleijn]{Kidess_2016_thermalSci}
		Kidess, A., Kenjere{\v{s}}, S., Righolt, B.~W., and Kleijn, C.~R.
		\newblock Marangoni driven turbulence in high energy surface melting processes.
		\newblock \emph{International Journal of Thermal Sciences}, 104:\penalty0
		412--422, 2016{\natexlab{a}}.
		\newblock \doi{10.1016/j.ijthermalsci.2016.01.015}.
		
		\bibitem[Grange \textit{et~al.}(2021)Grange, Queva, Guillemot, Bellet, Bartout,
		and Colin]{Grange_2021}
		Grange, D., Queva, A., Guillemot, G., Bellet, M., Bartout, J.-D., and Colin, C.
		\newblock Effect of processing parameters during the laser beam melting of
		{Inconel 738}: Comparison between simulated and experimental melt pool shape.
		\newblock \emph{Journal of Materials Processing Technology}, 289:\penalty0
		116897, 2021.
		\newblock \doi{10.1016/j.jmatprotec.2020.116897}.
		
		\bibitem[De and DebRoy(2004)]{De_2004}
		De, A. and DebRoy, T.
		\newblock A smart model to estimate effective thermal conductivity and
		viscosity in the weld pool.
		\newblock \emph{Journal of Applied Physics}, 95\penalty0 (9):\penalty0
		5230--5240, 2004.
		\newblock \doi{10.1063/1.1695593}.
		
		\bibitem[King \textit{et~al.}(2015)King, Anderson, Ferencz, Hodge, Kamath,
		Khairallah, and Rubenchik]{King_2015}
		King, W.~E., Anderson, A.~T., Ferencz, R.~M., Hodge, N.~E., Kamath, C.,
		Khairallah, S.~A., and Rubenchik, A.~M.
		\newblock Laser powder bed fusion additive manufacturing of metals; physics,
		computational, and materials challenges.
		\newblock \emph{Applied Physics Reviews}, 2\penalty0 (4):\penalty0 041304,
		2015.
		\newblock \doi{10.1063/1.4937809}.
		
		\bibitem[Khairallah and Anderson(2014)]{Khairallah_2014}
		Khairallah, S.~A. and Anderson, A.
		\newblock Mesoscopic simulation model of selective laser melting of stainless
		steel powder.
		\newblock \emph{Journal of Materials Processing Technology}, 214\penalty0
		(11):\penalty0 2627--2636, 2014.
		\newblock \doi{10.1016/j.jmatprotec.2014.06.001}.
		
		\bibitem[Indhu \textit{et~al.}(2018)Indhu, Vivek, Sarathkumar, Bharatish, and
		Soundarapandian]{Indhu_2018}
		Indhu, R., Vivek, V., Sarathkumar, L., Bharatish, A., and Soundarapandian, S.
		\newblock Overview of laser absorptivity measurement techniques for material
		processing.
		\newblock \emph{Lasers in Manufacturing and Materials Processing}, 5\penalty0
		(4):\penalty0 458--481, 2018.
		\newblock \doi{10.1007/s40516-018-0075-1}.
		
		\bibitem[Ye \textit{et~al.}(2019)Ye, Khairallah, Rubenchik, Crumb, Guss, Belak,
		and Matthews]{Ye_2019}
		Ye, J., Khairallah, S.~A., Rubenchik, A.~M., Crumb, M.~F., Guss, G., Belak, J.,
		and Matthews, M.~J.
		\newblock Energy coupling mechanisms and scaling behavior associated with laser
		powder bed fusion additive manufacturing.
		\newblock \emph{Advanced Engineering Materials}, 21\penalty0 (7):\penalty0
		1900185, 2019.
		\newblock \doi{10.1002/adem.201900185}.
		
		\bibitem[Svetlizky \textit{et~al.}(2021)Svetlizky, Das, Zheng, Vyatskikh, Bose,
		Bandyopadhyay, Schoenung, Lavernia, and Eliaz]{Svetlizky_2021}
		Svetlizky, D., Das, M., Zheng, B., Vyatskikh, A.~L., Bose, S., Bandyopadhyay,
		A., Schoenung, J.~M., Lavernia, E.~J., and Eliaz, N.
		\newblock Directed energy deposition ({DED}) additive manufacturing: Physical
		characteristics, defects, challenges and applications.
		\newblock \emph{Materials Today}, 2021.
		\newblock \doi{10.1016/j.mattod.2021.03.020}.
		
		\bibitem[Trapp \textit{et~al.}(2017)Trapp, Rubenchik, Guss, and
		Matthews]{Trapp_2017}
		Trapp, J., Rubenchik, A.~M., Guss, G., and Matthews, M.~J.
		\newblock In situ absorptivity measurements of metallic powders during laser
		powder-bed fusion additive manufacturing.
		\newblock \emph{Applied Materials Today}, 9:\penalty0 341--349, 2017.
		\newblock \doi{10.1016/j.apmt.2017.08.006}.
		
		\bibitem[Ready(1997)]{Ready_1997}
		Ready, J.
		\newblock \emph{Industrial applications of lasers}.
		\newblock Academic Press, San Diego, 2nd edition, 1997.
		\newblock ISBN 9780125839617.
		
		\bibitem[Katayama(2013)]{Katayama_2013}
		Katayama, S.
		\newblock \emph{Handbook of laser welding technologies}.
		\newblock Woodhead Publishing Limited, Philadelphia, PA, 2013.
		\newblock ISBN 9780857092649.
		
		\bibitem[Kouraytem \textit{et~al.}(2019)Kouraytem, Li, Cunningham, Zhao, Parab,
		Sun, Rollett, Spear, and Tan]{Kouraytem_2019}
		Kouraytem, N., Li, X., Cunningham, R., Zhao, C., Parab, N., Sun, T., Rollett,
		A.~D., Spear, A.~D., and Tan, W.
		\newblock Effect of laser-matter interaction on molten pool flow and keyhole
		dynamics.
		\newblock \emph{Physical Review Applied}, 11\penalty0 (6):\penalty0 064054,
		2019.
		\newblock \doi{10.1103/physrevapplied.11.064054}.
		
		\bibitem[Lvovsky(2015)]{Lvovsky_2015}
		Lvovsky, A.~I.
		\newblock Fresnel equations.
		\newblock In \emph{Encyclopedia of Optical and Photonic Engineering}. CRC
		Press, Boca Raton, Florida, 2nd edition, 2015.
		\newblock ISBN 9781351247184.
		
		\bibitem[Bass(1983)]{Bass_1983}
		Bass, M.
		\newblock \emph{Laser materials processing}.
		\newblock Elsevier Science, Amsterdam, North-Holland, 1983.
		\newblock ISBN 9780444863966.
		
		\bibitem[Simonds \textit{et~al.}(2018)Simonds, Sowards, Hadler, Pfeif, Wilthan,
		Tanner, Harris, Williams, and Lehman]{Simonds_2018}
		Simonds, B.~J., Sowards, J., Hadler, J., Pfeif, E., Wilthan, B., Tanner, J.,
		Harris, C., Williams, P., and Lehman, J.
		\newblock Time-resolved absorptance and melt pool dynamics during intense laser
		irradiation of a metal.
		\newblock \emph{Physical Review Applied}, 10\penalty0 (4):\penalty0 044061,
		2018.
		\newblock \doi{10.1103/physrevapplied.10.044061}.
		
		\bibitem[Ujihara(1972)]{Ujihara_1972}
		Ujihara, K.
		\newblock Reflectivity of metals at high temperatures.
		\newblock \emph{Journal of Applied Physics}, 43\penalty0 (5):\penalty0
		2376--2383, 1972.
		\newblock \doi{10.1063/1.1661506}.
		
		\bibitem[Wang and Yan(2021)]{Wang_2021}
		Wang, L. and Yan, W.
		\newblock Thermoelectric magnetohydrodynamic model for laser-based metal
		additive manufacturing.
		\newblock \emph{Physical Review Applied}, 15\penalty0 (6):\penalty0 064051,
		2021.
		\newblock \doi{10.1103/physrevapplied.15.064051}.
		
		\bibitem[Saldi(2012)]{Saldi_2012_thesis}
		Saldi, Z.~S.
		\newblock \emph{Marangoni driven free surface flows in liquid weld pools}.
		\newblock {PhD} dissertation, Delft University of Technology, Delft University
		of Technology, 2012.
		
		\bibitem[Jaques(1988)]{Jaques_1988}
		Jaques, A.
		\newblock Thermophysical properties of argon.
		\newblock Technical Report FNAL-TM-1517, Illinois, United States, 1988.
		
		\bibitem[Mills(2002)]{Mills_2002_316}
		Mills, K.~C.
		\newblock {Fe-316} stainless steel.
		\newblock In \emph{Recommended Values of Thermophysical Properties for Selected
			Commercial Alloys}, pages 135--142. Elsevier, 2002.
		\newblock \doi{10.1533/9781845690144.135}.
		
		\bibitem[Heeling \textit{et~al.}(2017)Heeling, Cloots, and
		Wegener]{Heeling_2017}
		Heeling, T., Cloots, M., and Wegener, K.
		\newblock Melt pool simulation for the evaluation of process parameters in
		selective laser melting.
		\newblock \emph{Additive Manufacturing}, 14:\penalty0 116--125, 2017.
		\newblock \doi{10.1016/j.addma.2017.02.003}.
		
		\bibitem[Khairallah \textit{et~al.}(2016)Khairallah, Anderson, Rubenchik, and
		King]{Khairallah_2016}
		Khairallah, S.~A., Anderson, A.~T., Rubenchik, A., and King, W.~E.
		\newblock Laser powder-bed fusion additive manufacturing: Physics of complex
		melt flow and formation mechanisms of pores, spatter, and denudation zones.
		\newblock \emph{Acta Materialia}, 108:\penalty0 36--45, 2016.
		\newblock \doi{10.1016/j.actamat.2016.02.014}.
		
		\bibitem[Kim(1975)]{Kim_1975}
		Kim, C.~S.
		\newblock Thermophysical properties of stainless steels.
		\newblock Technical Report ANL-75-55, Illinois, United States, 1975.
		
		\bibitem[Pichler \textit{et~al.}(2019)Pichler, Simonds, Sowards, and
		Pottlacher]{Pichler_2019}
		Pichler, P., Simonds, B.~J., Sowards, J.~W., and Pottlacher, G.
		\newblock Measurements of thermophysical properties of solid and liquid {NIST}
		{SRM} {316L} stainless steel.
		\newblock \emph{Journal of Materials Science}, 55\penalty0 (9):\penalty0
		4081--4093, 2019.
		\newblock \doi{10.1007/s10853-019-04261-6}.
		
		\bibitem[Sahoo \textit{et~al.}(1988)Sahoo, Debroy, and McNallan]{Sahoo_1988}
		Sahoo, P., Debroy, T., and McNallan, M.~J.
		\newblock Surface tension of binary metal{\textemdash}surface active solute
		systems under conditions relevant to welding metallurgy.
		\newblock \emph{Metallurgical Transactions B}, 19\penalty0 (3):\penalty0
		483--491, 1988.
		\newblock \doi{10.1007/bf02657748}.
		
		\bibitem[Hirt and Nichols(1981)]{Hirt_1981}
		Hirt, C.~W. and Nichols, B.~D.
		\newblock Volume of fluid ({VOF}) method for the dynamics of free boundaries.
		\newblock \emph{Journal of Computational Physics}, 39\penalty0 (1):\penalty0
		201--225, 1981.
		\newblock \doi{10.1016/0021-9991(81)90145-5}.
		
		\bibitem[Voller and Swaminathan(1991)]{Voller_1991}
		Voller, V.~R. and Swaminathan, C.~R.
		\newblock General source-based method for solidification phase change.
		\newblock \emph{Numerical Heat Transfer, Part B: Fundamentals}, 19\penalty0
		(2):\penalty0 175--189, 1991.
		\newblock \doi{10.1080/10407799108944962}.
		
		\bibitem[Voller and Prakash(1987)]{Voller_1987}
		Voller, V.~R. and Prakash, C.
		\newblock A fixed grid numerical modelling methodology for convection-diffusion
		mushy region phase-change problems.
		\newblock \emph{International Journal of Heat and Mass Transfer}, 30\penalty0
		(8):\penalty0 1709--1719, 1987.
		\newblock \doi{10.1016/0017-9310(87)90317-6}.
		
		\bibitem[Ebrahimi \textit{et~al.}(2019{\natexlab{a}})Ebrahimi, Kleijn, and
		Richardson]{Ebrahimi_2019}
		Ebrahimi, A., Kleijn, C.~R., and Richardson, I.~M.
		\newblock Sensitivity of numerical predictions to the permeability coefficient
		in simulations of melting and solidification using the enthalpy-porosity
		method.
		\newblock \emph{Energies}, 12\penalty0 (22):\penalty0 4360, 2019{\natexlab{a}}.
		\newblock \doi{10.3390/en12224360}.
		
		\bibitem[Brackbill \textit{et~al.}(1992)Brackbill, Kothe, and
		Zemach]{Brackbill_1992}
		Brackbill, J.~U., Kothe, D.~B., and Zemach, C.
		\newblock A continuum method for modeling surface tension.
		\newblock \emph{Journal of Computational Physics}, 100\penalty0 (2):\penalty0
		335--354, 1992.
		\newblock \doi{10.1016/0021-9991(92)90240-y}.
		
		\bibitem[Anisimov(1995)]{Anisimov_1995}
		Anisimov, S.~I.
		\newblock \emph{Instabilities in {Laser}-matter interaction}.
		\newblock CRC Press, Boca Raton, Fla, 1995.
		\newblock ISBN 0849386608.
		
		\bibitem[Lee \textit{et~al.}(2002)Lee, Ko, Farson, and Yoo]{Lee_2002}
		Lee, J.~Y., Ko, S.~H., Farson, D.~F., and Yoo, C.~D.
		\newblock Mechanism of keyhole formation and stability in stationary laser
		welding.
		\newblock \emph{Journal of Physics D: Applied Physics}, 35\penalty0
		(13):\penalty0 1570--1576, 2002.
		\newblock \doi{10.1088/0022-3727/35/13/320}.
		
		\bibitem[Johnson \textit{et~al.}(2017)Johnson, Rodgers, Underwood, Madison,
		Ford, Whetten, Dagel, and Bishop]{Johnson_2017}
		Johnson, K.~L., Rodgers, T.~M., Underwood, O.~D., Madison, J.~D., Ford, K.~R.,
		Whetten, S.~R., Dagel, D.~J., and Bishop, J.~E.
		\newblock Simulation and experimental comparison of the thermo-mechanical
		history and {3D} microstructure evolution of {304L} stainless steel tubes
		manufactured using {LENS}.
		\newblock \emph{Computational Mechanics}, 61\penalty0 (5):\penalty0 559--574,
		2017.
		\newblock \doi{10.1007/s00466-017-1516-y}.
		
		\bibitem[Sridharan \textit{et~al.}(2011)Sridharan, Allen, Anderson, Cao, and
		Kulcinski]{Sridharan_2011}
		Sridharan, K., Allen, T., Anderson, M., Cao, G., and Kulcinski, G.
		\newblock Emissivity of candidate materials for {VHTR} applicationbs: Role of
		oxidation and surface modification treatments.
		\newblock Technical report, University of Wisconsin, 2011.
		\newblock URL \url{https://www.osti.gov/biblio/1022709}.
		
		\bibitem[Ducharme \textit{et~al.}(1994)Ducharme, Williams, Kapadia, Dowden,
		Steen, and Glowacki]{Ducharme_1994}
		Ducharme, R., Williams, K., Kapadia, P., Dowden, J., Steen, B., and Glowacki,
		M.
		\newblock The laser welding of thin metal sheets: an integrated keyhole and
		weld pool model with supporting experiments.
		\newblock \emph{Journal of Physics D: Applied Physics}, 27\penalty0
		(8):\penalty0 1619--1627, 1994.
		\newblock \doi{10.1088/0022-3727/27/8/006}.
		
		\bibitem[Wooten(1972)]{Wooten_1972}
		Wooten, F.
		\newblock \emph{Optical properties of solids}.
		\newblock Academic Press, New York, 1972.
		\newblock ISBN 1483220761.
		
		\bibitem[{ANS}()]{Ansys_192}
		\emph{Release 19.2}.
		\newblock {ANSYS Fluent}.
		\newblock URL \url{https://www.ansys.com/}.
		
		\bibitem[Ebrahimi \textit{et~al.}(2019{\natexlab{b}})Ebrahimi, Kleijn, and
		Richardson]{Ebrahimi_2019_conf}
		Ebrahimi, A., Kleijn, C.~R., and Richardson, I.~M.
		\newblock The influence of surface deformation on thermocapillary flow
		instabilities in low {Prandtl} melting pools with surfactants.
		\newblock In \emph{{Proceedings of the 5th World Congress on Mechanical,
				Chemical, and Material Engineering}}. Avestia Publishing, 2019{\natexlab{b}}.
		\newblock \doi{10.11159/htff19.201}.
		
		\bibitem[Ebrahimi \textit{et~al.}(2021{\natexlab{d}})Ebrahimi, Babu, Kleijn,
		Hermans, and Richardson]{Ebrahimi_2021_c}
		Ebrahimi, A., Babu, A., Kleijn, C.~R., Hermans, M. J.~M., and Richardson, I.~M.
		\newblock The effect of groove shape on molten metal flow behaviour in gas
		metal arc welding.
		\newblock \emph{Materials}, 14\penalty0 (23):\penalty0 7444,
		2021{\natexlab{d}}.
		\newblock \doi{10.3390/ma14237444}.
		
		\bibitem[Patankar(1980)]{Patankar_1980}
		Patankar, S.~V.
		\newblock \emph{Numerical Heat Transfer and Fluid Flow}.
		\newblock Taylor \& Francis Inc, 1\textsuperscript{st} edition, 1980.
		\newblock ISBN 0891165223.
		
		\bibitem[Issa(1986)]{Issa_1986}
		Issa, R.~I.
		\newblock Solution of the implicitly discretised fluid flow equations by
		operator-splitting.
		\newblock \emph{Journal of Computational Physics}, 62\penalty0 (1):\penalty0
		40--65, 1986.
		\newblock \doi{10.1016/0021-9991(86)90099-9}.
		
		\bibitem[Ubbink(1997)]{Ubbink_1997}
		Ubbink, O.
		\newblock \emph{Numerical Prediction of Two Fluid Systems with Sharp
			Interfaces}.
		\newblock {PhD} dissertation, Imperial College London (University of London),
		London, United Kingdom, 1997.
		\newblock URL \url{http://hdl.handle.net/10044/1/8604}.
		
		\bibitem[Kell \textit{et~al.}(2006)Kell, Tyrer, Higginson, Thomson, Jones, and
		Noden]{Kell_2006}
		Kell, J., Tyrer, J., Higginson, R., Thomson, R., Jones, J., and Noden, S.
		\newblock Holographic diffractive optical elements allow improvements in
		conduction laser welding of steels.
		\newblock In \emph{{International Congress on Applications of Lasers {\&}
				Electro-Optics}}. Laser Institute of America, 2006.
		\newblock \doi{10.2351/1.5060749}.
		
		\bibitem[Mills \textit{et~al.}(1998)Mills, Keene, Brooks, and
		Shirali]{Mills_1998}
		Mills, K.~C., Keene, B.~J., Brooks, R.~F., and Shirali, A.
		\newblock Marangoni effects in welding.
		\newblock \emph{Philosophical Transactions of the Royal Society A:
			Mathematical, Physical and Engineering Sciences}, 356\penalty0
		(1739):\penalty0 911--925, 1998.
		\newblock \doi{10.1098/rsta.1998.0196}.
		
		\bibitem[Kidess \textit{et~al.}(2016{\natexlab{b}})Kidess, Kenjere{\v{s}}, and
		Kleijn]{Kidess_2016_PhysFlu}
		Kidess, A., Kenjere{\v{s}}, S., and Kleijn, C.~R.
		\newblock The influence of surfactants on thermocapillary flow instabilities in
		low {Prandtl} melting pools.
		\newblock \emph{Physics of Fluids}, 28\penalty0 (6):\penalty0 062106,
		2016{\natexlab{b}}.
		\newblock \doi{10.1063/1.4953797}.
		
		\bibitem[Zhao \textit{et~al.}(2010)Zhao, Kwakernaak, Pan, Richardson, Saldi,
		Kenjeres, and Kleijn]{Zhao_2010}
		Zhao, C.~X., Kwakernaak, C., Pan, Y., Richardson, I.~M., Saldi, Z., Kenjeres,
		S., and Kleijn, C.~R.
		\newblock The effect of oxygen on transitional {Marangoni} flow in laser spot
		welding.
		\newblock \emph{Acta Materialia}, 58\penalty0 (19):\penalty0 6345--6357, 2010.
		\newblock \doi{10.1016/j.actamat.2010.07.056}.
		
		\bibitem[Schatz and Neitzel(2001)]{Schatz_2001}
		Schatz, M.~F. and Neitzel, G.~P.
		\newblock Experiments on thermocapillary instabilities.
		\newblock \emph{Annual Review of Fluid Mechanics}, 33\penalty0 (1):\penalty0
		93--127, 2001.
		\newblock \doi{10.1146/annurev.fluid.33.1.93}.
		
		\bibitem[Nakamura \textit{et~al.}(2015)Nakamura, Kawahito, Nishimoto, and
		Katayama]{Nakamura_2015}
		Nakamura, H., Kawahito, Y., Nishimoto, K., and Katayama, S.
		\newblock Elucidation of melt flows and spatter formation mechanisms during
		high power laser welding of pure titanium.
		\newblock \emph{Journal of Laser Applications}, 27\penalty0 (3):\penalty0
		032012, 2015.
		\newblock \doi{10.2351/1.4922383}.
		
	\end{thebibliography}

\end{document}